\documentclass[11pt,a4paper]{article}
\usepackage{jstylemarg}
\usepackage{longtable}
\usepackage[vcentermath]{youngtab}

\newcommand{\bi}{\begin{itemize}}
\newcommand{\ei}{\end{itemize}}
\newcommand{\bea}{\begin{align}}
\newcommand{\eea}{\end{align}}
\newcommand{\be}{\begin{equation}}
\newcommand{\ee}{\end{equation}}
\newcommand{\m}[1]{$\mathop{#1}$}
\newcommand{\la}{\langle}
\newcommand{\ra}{\rangle}
\newcommand{\ba}{\begin{eqnarray}}
\newcommand{\ea}{\end{eqnarray}}

\newcommand{\sst}{\scriptscriptstyle}

\newcommand{\nn}{\nonumber\\}

\usepackage{mathtools}

\newcommand*\circled[1]{\tikz[baseline=(char.base)]{
  \node[shape=circle,draw, inner sep=1pt] (char) {#1};}}
\newcommand*\dcircled[1]{\tikz[baseline=(char.base)]{
  \node[shape=circle,fill=black!30, draw,inner sep=1pt] (char) {#1};}}

\newcommand{\gad}{{\dot{\alpha}}}

\newcommand{\ga}{\alpha}

\newcommand{\gA}{\mathrm{A}}
\newcommand{\gB}{\mathrm{B}}
\newcommand{\gC}{\mathrm{C}}
\newcommand{\gD}{\mathrm{D}}

\newcommand{\Tr}{{\text{Tr}}}
\newcommand{\Det}{{\text{det}}}
\newcommand{\pl}{{\partial}}

\newcommand{\bry}{{{\bar{y}}}}

\newcommand{\tcb}{\textcolor{blue}}

\makeatletter
\renewcommand*\env@matrix[1][\arraystretch]{%
  \edef\arraystretch{#1}%
  \hskip -\arraycolsep
  \let\@ifnextchar\new@ifnextchar
  \array{*\c@MaxMatrixCols c}}
\makeatother

\author{Charlotte SLEIGHT$^{a}$}

\author{\quad Massimo TARONNA$^{b}$}

\affiliation{${}^{a}$Max-Planck-Institut f\"ur Physik\\
\hspace*{0.01cm} F\"ohringer Ring 6, 80805 Munich, Germany}

\affiliation{${}^{b}$Universit\'e Libre de Bruxelles
and International Solvay Institutes\\
\hspace*{0.01cm} ULB-Campus Plaine CP231, 1050 Brussels, Belgium}

\emailAdd{charlotte.sleight@gmail.com, massimo.taronna@ulb.ac.be}

\title{\centering
\huge{Higher-Spin Algebras, Holography\\ and Flat Space}
}

\abstract{In this article we study the higher-spin algebra behind the type-A cubic couplings recently extracted from the free $O(N)$ model in generic dimensions, demonstrating that they coincide with the known structure constants for the unique higher-spin algebra in generic dimensions. This provides an explicit check of the holographic reconstruction and of the duality between higher-spin theories and the free $O(N)$ model in generic dimensions, generalising the result of Giombi and Yin in AdS$_4$. For completeness, we also address the same problem in the flat space for the cubic couplings derived by Metsaev in 1991, which are recovered from the flat limit of the AdS type-A cubic couplings. We observe that both flat and AdS$_4$ higher-spin Lorentz subalgebras coincide, hinting towards the existence of a full higher-spin symmetry behind the flat-space cubic couplings of Metsaev.}

\begin{document}
\begin{flushright}    
  {MPP-2016-274}
\end{flushright}
\maketitle

\section{Introduction}

In this paper we investigate the algebras underlying higher-spin theories on AdS- and flat-space, as implied by the available expressions for the metric-like action at cubic order.

Recently, a cubic order action for the type A minimal higher-spin theory on AdS$_{d+1}$ was obtained employing its conjectured duality \cite{Sezgin:2002rt,Klebanov:2002ja} with the free scalar $O\left(N\right)$ model \cite{Sleight:2016dba,Sleight:2016hyl}.\footnote{See \cite{Petkou:2003zz,Bekaert:2015tva,Skvortsov:2015lja,Sleight:2016hyl} for preliminary works on the holographic reconstruction of the $0$-$0$-$s$ couplings. At quartic order, see \cite{Bekaert:2014cea,Bekaert:2015tva} for the reconstruction of the quartic self-interaction of the scalar. See also \cite{Skvortsov:2015pea} for the $0$-$0$-$s$ coupling in the type B theory.} A non-trivial test of this duality would then be provided from the comparison of the rigid structure constants implied by the holographically reconstructed cubic action with the known expressions \cite{Vasiliev:2003ev,Joung:2014qya} for higher-spin  algebra structure constants, which are unique in general dimensions \cite{Boulanger:2013zza}. This check of the duality is investigated in this work, with the outcome that the result is indeed positive, and with it extending to general dimensions the three-point function test \cite{Giombi:2009wh} of Giombi and Yin in AdS$_4$.\footnote{See also \cite{Chang:2011mz,Ammon:2011ua} for subsequent tree-level three point function tests in AdS$_3$.} In particular, the test confirms that the holographically reconstructed cubic couplings solve the Noether procedure at the quartic order.  

Re-winding a number of years, in \cite{Metsaev:1991mt,Metsaev:1991nb} Metsaev constructed the complete cubic action for the higher-spin theory in flat space using light-cone methods. Motivated by the observation that it contains vertices which are not accommodated for in the original covariant classification \cite{Boulanger:2008tg,Zinoviev:2008ck,Manvelyan:2010jr,Sagnotti:2010at,Fotopoulos:2010ay} of cubic vertices in flat space, we study the corresponding rigid structure constants. This is particularly appealing, for the additional vertices in Metsaev's solution are lower derivative, including a two-derivative coupling of higher-spin fields with gravity \cite{Bengtsson:1983pd,Bengtsson:1986kh}. We argue that this coupling can be considered as a minimal coupling, in accordance with the equivalence principle.\footnote{That the cubic two-derivative couplings of any soft particle must be equal was shown by Weinberg in his seminal paper \cite{Weinberg:1964ew}. As we review in \S \tcb{\ref{quarticorder}}, his conclusion must also hold even in the case of trivial S-matrix amplitudes. A similar discussion was also made in \cite{Metsaev:1991mt} by Metsaev. There, upon recovering charge conservation and equivalence principle directly in the light cone gauge for spin-1 and spin-2 fields, a unique class of  solutions for higher-spin fields coupled to gravity is found in four-dimensions which indeed obey the equivalence principle and verifies the equality of all higher-spin two-derivative gravitational couplings. Furthermore, Metsaev's solution also gives a generalisation of this result to the lowest derivative $s$-$s$-$s^\prime$ couplings, which turn out to be equal for all $s$ with the only dependence being on $s^\prime$.}  Indeed, in \cite{Joung:2013nma} a version of the Coleman-Mandula theorem was proven, stating that the flat space cubic interactions in the original covariant classification cannot give rise to a higher-spin theory with higher-spin generators satisfying non-trivial commutation relations with the isometry generators.\footnote{See also \cite{Bekaert:2010hp}, and \cite{Bekaert:2010hw} for a nice review of no-go theorems in the context of higher-spins.} We extract explicit expressions for the Lorentz part of the structure constants, which we find matches with the Lorentz part of corresponding AdS$_4$ higher-spin algebra. The existence of these structure constants crucially relies on the presence of the additional lower derivative vertices, which are local in the light-cone gauge but (as we shall demonstrate) do not admit a standard covariant form. Of course, in spite of these promising results it remains to be seen whether other consistency conditions (such as those from higher-orders in the Noether procedure) permit the existence of a consistent unitary interacting theory. See \cite{Bengtsson:2016jfk,Ponomarev:2016jqk,Bengtsson:2016alt,Bengtsson:2016hss,Ponomarev:2016lrm,Taronna:2017wbx} for recent works in this direction.

\subsection{Outline}

The article is organised as follows: In \S \tcb{\ref{AdS}} we review the deformation of gauge transformations by higher-spin interactions and extract the rigid structure constants from the cubic vertices of \cite{Sleight:2016dba}. \S \tcb{\ref{Metsaev}} is devoted to the $4d$ cubic action of Metsaev \cite{Metsaev:1991mt,Metsaev:1991nb} and to extracting the corresponding structure constants upon considering a formal covariantisation of the light-cone vertices. We then argue that these structure constants give a well-defined higher-spin symmetry, and that the two-derivative coupling of a higher-spin field to gravity can indeed be considered as a minimal coupling.
In \S \tcb{\ref{Sec:spinohelicity}} we consider the aforementioned flat space cubic vertices in the framework of the spinor-helicity formalism. In \S \tcb{\ref{quarticorder}} we discuss some consequences of such a higher-spin symmetry on higher order amplitudes. Concluding \S \tcb{\ref{Conclusions}} summarises the results of the paper and presents few outlooks. \S \tcb{\ref{appendix}} gives a summary of higher-spin algebra structure constants.

The methods we employ to extract the structure constants were developed in previous works \cite{Joung:2011ww,Joung:2012rv,Joung:2012fv,Taronna:2012gb,Joung:2013doa,Joung:2013nma}, which we briefly review in this paper. In particular, we extract the structure constants from the first-order deformations of the gauge transformations induced by cubic interactions. This is reviewed in the following section.

\subsection{Review: Noether approach to higher-spin theories} 
\label{subsec::noerev}
The Noether method to constructing interacting higher-spin gauge theories is a systematic perturbative approach, underpinned by the requirement of gauge invariance \cite{Berends:1984rq}: For a given spectrum, one begins with the free theory and adds interactions order by order in the weak fields in a way that is consistent with the gauge symmetries at each order. 

From an action perspective, this reads
\begin{align}
S &= S^{\left(2\right)}+S^{\left(3\right)} + ... \, ,& \quad \delta_{\xi}  &=  \delta^{\left(0\right)}_{\xi}  + \delta^{\left(1\right)}_{\xi} + ... \, ,
\end{align}
where the superscript $\left(n\right)$ indicates that the corresponding term involves $n$ powers of the fields. The condition of gauge invariance is then translated into an infinite set of coupled equations,
\be
	\delta_{\xi}\,S=0
	\quad\Rightarrow\quad
	\left\{\begin{array}{cc}
	\delta^{\sst (0)}_{\xi}\,S^{\sst (2)}=0 &\quad \circled{\scriptsize 0}\\
	\delta^{\sst (0)}_{\xi}\,S^{\sst (3)}
	+\delta^{\sst (1)}_{\xi}\,S^{\sst (2)}=0 & 
	\quad  \circled{\scriptsize 1}\\
	\delta^{\sst (0)}_{\xi}\,S^{\sst (4)}
	+\delta^{\sst (1)}_{\xi}\,S^{\sst (3)}
	+\delta^{\sst (2)}_{\xi}\,S^{\sst (2)}=0 & \quad 
	\circled{\scriptsize 2}\\
	\vdots
	\end{array}
	\right.,
	\label{gauge inv}
\ee
Given the free action $S^{\sst (2)}$ and corresponding gauge transformations $\delta^{\sst (0)}_{\xi}\varphi$, interaction terms in the action $S^{\sst (n\ge3)}$\, and the higher-order gauge transformations $\delta^{\sst (n\ge1)}_{\xi}\varphi$\, can be determined by solving the system \eqref{gauge inv} iteratively under certain locality assumptions (see \cite{Barnich:1993vg,Bekaert:2010hp,Taronna:2011kt,Vasiliev:2015wma,Vasiliev:2015mka,Kessel:2015kna,Boulanger:2015ova,Bekaert:2015tva,Skvortsov:2015lja,Taronna:2016ats,Bekaert:2016ezc,Taronna:2016xrm,Sleight:2016hyl,Taronna:2017wbx,Roiban:2017iqg} for various discussions):
\be
	S^{\sst (2)}\,,\, \delta^{\sst (0)}_{\xi}\varphi
	\ \ \overset{\dcircled{\tiny 1}}{\bm\longrightarrow} \ \ 
	S^{\sst (3)}
	\ \ \overset{\circled{\tiny 1}}{\bm\longrightarrow} \ \
	\delta^{\sst (1)}_{\xi}\varphi
	\ \ \overset{\dcircled{\tiny 2}}{\bm\longrightarrow} \ \
	S^{\sst (4)}
	\ \ \overset{\circled{\tiny 2}}{\bm\longrightarrow} \ \
	\delta^{\sst (2)}_{\xi}\varphi
	\ \ {\bm\longrightarrow} \ \
	\cdots\,,
\ee
where $\dcircled{\tiny n}$ represents the same condition as $\circled{\tiny n}$ but solved this time
on the shell of \emph{free} EoM. 
In particular, at each order one can first solve for 
$S^{\sst (n+2)}$ using  $\dcircled{\tiny n}$
and then read off $\delta^{\sst (n)}_{\xi}\varphi$
from $\circled{\tiny n}$\,. 

The fully non-linear gauge transformations must also form an (open) algebra, which is the requirement\footnote{This is a necessary condition for the gauge orbits in field space to be integrable, i.e. that the infinitesimal transformation originates from a finite one.} \cite{Berends:1984rq}  
\be
\delta_{\xi_1} \delta_{\xi_2} -  \delta_{\xi_2} \delta_{\xi_1} = \delta_{\left[\!\right[\xi_1,\xi_2\left]\!\right]} \:+\:\text{on-shell trivial}.
\ee
Perturbatively, this is
\be
\delta^{\left(0\right)}_{\xi_1} \delta^{\left(n\right)}_{\xi_2} +\delta^{\left(1\right)}_{\xi_1} \delta^{\left(n-1\right)}_{\xi_2} + \: ... \: + \delta^{\left(n-1\right)}_{\xi_1} \delta^{\left(1\right)}_{\xi_2} \: - \: \left(\xi_1 \leftrightarrow \xi_2\right) =  \delta^{\left(0\right)}_{\left[\!\right[\xi_1,\xi_2\left]\!\right]^{\left(n-1\right)}} + \: ... \: +  \delta^{\left(n-1\right)}_{\left[\!\right[\xi_1,\xi_2\left]\!\right]^{\left(0\right)}}, \label{closure}
\ee
where we have also expanded the field-dependent commutator
\be
\left[\!\right[\xi_1,\xi_2\left]\!\right] = \left[\!\right[\xi_1,\xi_2\left]\!\right]^{\left(0\right)}+\left[\!\right[\xi_1,\xi_2\left]\!\right]^{\left(1\right)} + \: ...\;.\label{comexp}
\ee
From the lowest order condition \eqref{closure}
\be
\delta^{\left(0\right)}_{\xi_1} \delta^{\left(1\right)}_{\xi_2} - \delta^{\left(0\right)}_{\xi_2} \delta^{\left(1\right)}_{\xi_1} =  \delta^{\left(0\right)}_{\left[\!\right[\xi_1,\xi_2\left]\!\right]^{\left(0\right)}}, \label{closure0}
\ee
the field-independent part of the commutator \eqref{comexp} can be identified from $\delta^{\left(1\right)}_{\xi}$. 

A further crucial requirement is that the rigid (global) symmetry associated to the commutator \eqref{comexp} defines a Lie algebra, i.e. that it satisfies the Jacobi identity. This is consequence of associativity of gauge transformations and closure \eqref{closure} at second order ($n=2$). The rigid commutators derive from the lowest order commutator $\left[\!\right[\xi_1,\xi_2\left]\!\right]^{\left(0\right)}$, evaluated on solutions $\xi = {\bar \xi}$ to the Killing equations:
\be
0 = \left[\, \delta_\xi \varphi_{\mu_1...\mu_s}\, \right]_{\varphi=0} = \nabla_{\left(\mu_1\right.}\xi_{\mu_2...\mu_s\left.\right)}.
\ee
In summary, in this paper we analyse this condition for the cubic action \cite{Sleight:2016dba} established by holographic reconstruction for the type A minimal theory on AdS$_{d+1}$. The latter cubic action was not obtained via the standard Noether approach \eqref{gauge inv}, and since higher-spin algebra structure constants are unique in generic dimensions, this study provides a non-trivial check of the holographic reconstruction and a test of the holographic duality. We also analyse this condition for the cubic couplings of the flat space theory \cite{Metsaev:1991mt,Metsaev:1991nb}, which was obtained by requiring closure of the Poincar\'e algebra on the light-cone. 

\section{The holographic cubic action and induced gauge symmetries}\label{AdS}

\subsection{Review: AdS cubic couplings}

In this section we review the construction of cubic interactions in higher-spin gauge theory on anti-de Sitter (AdS) space, in particular the completion of the theory at cubic order via holographic reconstruction \cite{Sleight:2016dba}. The relevant results are recalled up to terms proportional to divergences and traces of the gauge fields. This traceless and transverse (TT) framework is sufficient for the purpose of extracting the corresponding putative higher-spin algebra structure constants (see e.g. \cite{Joung:2013nma}), which will be reviewed in \S \ref{subsec:::defgaugesymm}. In the sequel all equalities therefore hold modulo terms proportional to traces and divergences, which we denote by $\overset{\text{TT}}{=}$ unless the context is clear. 

\subsubsection*{Ambient-space formalism}

The results for the cubic action are most conveniently expressed and obtained in the ambient space formulation (see \cite{Taronna:2012gb,Sleight:2017krf} for further details). In this framework, AdS$_{d+1}$ space with radius $R$ is realised as a hyperboloid in an ambient $\left(d+2\right)$-dimensional Minkowski space,
\begin{equation}
    X^2+R^2=0, \qquad X^0 > 0.
\end{equation}
From this point onwards we set $R=1$. Symmetric spin-$s$ fields $\varphi_{\mu_1...\mu_s}$ intrinsic to the AdS manifold are described in this framework by ambient avatars $\Phi_{M_1 ... M_s}\left(X\right)$,\footnote{In particular, with pullback\begin{equation}
    \varphi_{\mu_1...\mu_s}\left(x\right) = \frac{\partial X^{M_1}\left(x\right)}{\partial x^{\mu_1}} ... \frac{\partial X^{M_s}\left(x\right)}{\partial x^{\mu_s}} \Phi\left(X\left(x\right)\right) \big|_{X^2=-R^2}.
\end{equation}} which satisfy homogeneity and tangentiality constraints
\begin{align}
 (X\cdot\pl_X-U \cdot \partial_U-\tau)\Phi(X,U)=0\,, \qquad X\cdot\pl_U\,\Phi(X,U)&=0.\,
\end{align}
In the above we introduced the generating function 
\begin{equation}
    \Phi\left(X,U\right) = \sum^\infty_{s=0} \frac{1}{s!} \Phi_{M_1 ... M_s}\left(X\right) U^{M_1} ... U^{M_s}\,,\label{gf}
\end{equation}
where $U^M$ is an ambient auxiliary vector. When $\tau = -2$, the generating function \eqref{gf} packages a tower of bosonic spin-$s$ gauge fields with gauge symmetries 
\begin{align}
    \delta_{E} \Phi(U)&=U\cdot\pl_XE(U) + {\cal O}\left(\Phi\right)\,, \label{gt}
\end{align}
providing an ambient description of the intrinsic gauge transformations
\begin{equation}
    \delta_{\xi} \varphi_{\mu_1...\mu_s} = \nabla_{\left(\mu_1\right.} \xi_{\left. \mu_2...\mu_s\right)}+{\cal O}\left(\varphi\right).
\end{equation}

\subsubsection*{Traceless and transverse cubic action}
\label{subsubsec::TT}
The first non-trivial consistency condition \eqref{gauge inv} (i.e.\,$\circled{\scriptsize 0}$) fixes the kinetic term of the higher-spin action. In the ambient space formalism, this reads\footnote{Notice that here we use an alternative normalisation to the preceding paper \cite{Sleight:2016dba}. The latter normalisation can be obtained from the present one by replacing $\Phi_{M_1...M_s} \rightarrow \sqrt{s!}\,\Phi_{M_1...M_s}$.}
\begin{equation}
    S^{(2)}=\frac{1}{2}\int_{\text{AdS}_{d+1}}\,e^{\pl_{U_1}\cdot\pl_{U_2}}\left[\Phi(U_1)\,\Box\Phi(U_2)+\ldots\,\right]_{U_i=0}\,,
\end{equation}
where the $\ldots$ are TT contributions which we disregard and for convenience we use a non-canonical normalisation.

\begin{figure}[t]
  \centering
  \includegraphics[scale=0.5]{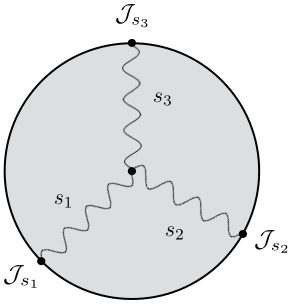} \label{3ptw}
  \caption{Holographic reconstruction of the bulk $s_1$-$s_2$-$s_3$ cubic vertex from the three-point function $\langle {\cal J}_{s_1}{\cal J}_{s_2}{\cal J}_{s_3} \rangle$ of single-trace operators of spins $s_i$ in the free scalar $O\left(N\right)$ model.} 
\end{figure}

At cubic order, the TT part of a coupling can be encoded modulo total derivatives by a function of six building blocks
\begin{subequations}\label{adsbb}
\begin{align}
Y_1&=\pl_{U_1}\cdot\pl_{X_2}\,,& Y_2&=\pl_{U_2}\cdot\pl_{X_3}\,,& Y_3&=\pl_{U_3}\cdot\pl_{X_1}\,,\\
 H_1&=\pl_{U_2}\cdot\pl_{U_3}\,,&  H_2&=\pl_{U_3}\cdot\pl_{U_1}\,,&  H_3&=\pl_{U_1}\cdot\pl_{U_2}\,.
\end{align}
\end{subequations}
The most general ansatz for the TT part of the cubic action can then be expressed in the form 
\begin{equation}
    S^{(3)}\overset{\text{TT}}{=}\int_{\text{AdS}_{d+1}}f(Y_i, H_i)\\\,\Phi_1\,\Phi_2\,\Phi_3\Big|_{U_i=0,X_i=X}\,,
\end{equation}
where the function $f(Y_i, H_i)$ may be fixed by enforcing Noether consistency \eqref{gauge inv}. The latter at step $\dcircled{\scriptsize 1}$ yields the constraint,
\begin{equation}
    \delta^{(0)}_{\xi}S^{(3)} \approx 0,
\end{equation}
whose solution restricts $f(Y_i, H_i)$ to the form \cite{Joung:2011ww,Joung:2013nma}
\begin{align}
f(Y_i, H_i) = \sum_{s_i,\,n} k^n_{s_1s_2s_3} P^{[n]}_{s_1s_2s_3}, \qquad P^{[n]}_{s_1s_2s_3} = e^{\mathfrak{D}} Y^{s_1-n}Y^{s_2-n}Y^{s_3-n}G^n, \label{frel}
\end{align}
where 
\begin{align}
   G&\equiv Y_1 H_1+ Y_2 H_2+ Y_3 H_3\,, \label{gads}
\end{align}
and $\mathfrak{D}$ is the differential operator\footnote{$\lambda$ arises from the use of the ambient space measure introduced in \cite{Joung:2011ww}, with
\begin{equation}\label{lambda}
    \lambda^n\equiv(-1)^n(\Delta+d)(\Delta+d-2)\ldots(\Delta+d-2n+2).
\end{equation}
Here $\Delta$ is the total degree of homogeneity of a given term in the vertex.}
\begin{align}
\mathfrak{D}&=\lambda\left( H_1\pl_{Y_2}\pl_{Y_3}+ H_2 H_3\pl_{Y_1}\pl_{ G_3}+\text{cycl.}\right)+\,\lambda\, H_1 H_2 H_3\,\pl_{ G}^2\,.\label{mfd}
\end{align}
In principle the coefficients $k^n_{s_1s_2s_3}$ in \eqref{frel} may be determined from the higher order consistency conditions in \eqref{gauge inv}. An alternative route was taken in \cite{Sleight:2016dba}, using the holographic duality between the type A minimal bosonic higher-spin theory and the free scalar $O\left(N\right)$ vector model. By matching the three-point Witten diagrams in the bulk theory to the dual CFT correlation functions of single-trace operators (figure \ref{3ptw}), the cubic vertices for any triplet of spin $\left\{s_1,s_2,s_3\right\}$ have been determined
\begin{subequations}
\begin{align}
    (f_{\text{TT}})_{s_1,s_2,s_3}&=g_{s_1,s_2,s_3}\,e^{\widetilde{\mathfrak{D}}}\,Y_1^{s_1}Y_1^{s_2}Y_1^{s_3}\,,\\
    \widetilde{\mathfrak{D}}&=\lambda\left( H_1\pl_{Y_2}\pl_{Y_3}-2\, H_2 H_3\pl_{Y_2}\pl_{Y_3}\pl_{Y_1}^2+\text{cycl.}\right)\\&\hspace{100pt}+4\,\lambda H_1 H_2 H_3\,\pl_{Y_1}^2\pl_{Y_2}^2\pl_{Y_3}^2-2\, G\,\pl_{Y_1}\pl_{Y_2}\pl_{Y_3}\,,\nonumber
\end{align}
\end{subequations}
with relative coupling constants\footnote{Note the extra factors of $\sqrt{s_i!}$ compared to \cite{Sleight:2016dba}, due to the different choice of kinetic term normalisation.}
\begin{align}
   g_{s_1,s_2,s_3}= \frac{1}{\sqrt{N}}\frac{\pi ^{\frac{d-3}{4}}2^{\tfrac{3 d-1+s_1+s_2+s_3}{2}}}{s_1!\,s_2!\,s_3!\, \Gamma (d+s_1+s_2+s_3-3)}\prod_{i=1}^3\sqrt{\Gamma(s_i+\tfrac{d-1}{2})}\,.
\end{align}
Assuming the holographic duality holds, the full TT cubic action thus reads
\begin{equation}\label{full}
 S^{(3)}\overset{\text{TT}}{=}\int_{\text{AdS}_{d+1}}\sum_{s_i}\,(f_{\text{TT}})_{s_1,s_2,s_3}\,\Phi_1(X_1,U_1)\,\Phi_2(X_2,U_2)\,\Phi_3(X_3,U_3) \Big|_{U_i=0,\,X_i=X}\,.
\end{equation}
From this result one may test the holographic duality by extracting the corresponding global symmetry structure constants and comparing with the known expressions \cite{Joung:2014qya}. This is carried out in the following sections.

\subsection{Deformation of the gauge symmetries}
\label{subsec:::defgaugesymm}

In order to preserve the number of degrees of freedom, introducing interactions induces deformations in the gauge transformations. Cubic interactions may lead to ${\cal O}\left(\varphi\right)$ deformations, which can be seen from the $\circled{\scriptsize 1}$ consistency condition in \eqref{gauge inv}
\begin{equation}
    \delta^{(1)}_{\xi} S^{(2)} + \delta^{(0)}_{\xi} S^{(3)} = 0.\label{con1}
\end{equation}
In this section we determine such corrections induced by the holographically reconstructed cubic action \eqref{full}. We employ the approach taken in \cite{Joung:2013nma}, which used ambient space techniques to extract the deformations necessitated by the cubic structures \eqref{frel}.

Given a cubic action, the idea is to extract the corresponding deformation of the gauge transformations from the consistency condition \eqref{con1}. The first step is to compute the variation of the cubic vertices off-shell under linearised gauge transformations. This is proportional to the equations of motion (by consistency), and for the holographic cubic action \eqref{full} this reads
\begin{equation}\label{deltaS3}
\delta_{E_1}^{(0)}S^{(3)}=\int_{\text{AdS}_{d+1}}\,\left[\pl_{Y_1}f_\text{TT}\right]\,E_1(X_1,U_1)\,\tfrac12\,(\pl_{X_3}^2-\pl_{X_2}^2)\Phi_2(X_2,U_2)\Phi_3(X_3,U_3)\,.
\end{equation}
To satisfy \eqref{con1}, the above must then be compensated by the variation of the quadratic part of the action
\begin{multline}\label{deltaS2}
\delta^{(1)}S^{(2)}_{E_1}=\int_{\text{AdS}_{d+1}}e^{\pl_{U_1}\cdot\pl_{U_2}}\Big\{\delta_{E_1}^{(1)}\Phi_2(X,U_1)\,\pl_X^2\Phi_2(X,U_2)\\+\delta_{E_1}^{(1)}\Phi_3(X,U_1)\,\pl_X^2\Phi_3(X,U_3)
\Big\}_{U_i=0}\,.
\end{multline}
The key to then solve \eqref{con1} for the  deformations is to rewrite \eqref{deltaS3} in a way that factorises the equations of motion, as in \eqref{deltaS2}. This can be achieved by simply integrating by parts, which leads to\footnote{To this end, it is convenient to switch from encoding the vertex with basis \eqref{adsbb} to the following  
\begin{align}
\bar{Y}_1 := \pl_{U_1}\cdot\pl_{X_2}\,, \qquad
\bar{Y}_2 := -\pl_{U_2}\cdot\pl_{X_1}\,,\qquad
\bar{Y}_3 := \pl_{U_3}\cdot\pl_{X_1}\,.
\end{align}.}
\begin{equation}
\delta_{E_1}^{(0)}S^{(3)}=\int_{\text{AdS}_{d+1}}\,e^{\pl_{U_1}\cdot\pl_{U_2}}\left[ T_{12}(X,U_1)\,\pl_X^2\Phi_3(X,U_2)+T_{13}(X,U_1)\,\pl_X^2\Phi_2(X,U_2)\right]\,.
\end{equation}
where we introduced
\begin{equation}
T_{12}(X,U_1)=\tfrac12\,\pl_{\bar{Y}_1}\,
\bar{f}(\bar{{Y}},{H})\,E_1(X_1,U_1)\,\Phi_2(X_2,U_2)\,,
\end{equation}
with
\begin{equation}
\bar{f}(\bar{Y},{H})=\sum_{s_i}g_{s_1,s_2,s_3}\,e^{\bar{\mathfrak{D}}}\,\bar{Y}_1^{s_1}\bar{Y}_2^{s_2}\bar{Y}_3^{s_3}\,,
\end{equation}
and
\begin{align}
\bar{\mathfrak{D}}&=\lambda\left( H_1\pl_{\bar{Y}_2}\pl_{\bar{Y}_3}+ H_2\pl_{\bar{Y}_3}\pl_{\bar{Y}_1}-2\, H_1 H_2\,\pl_{\bar{Y}_1}\pl_{\bar{Y}_2}\pl_{\bar{Y}_3}^2\right)-2\, G\,\pl_{\bar{Y}_1}\,\pl_{\bar{Y}_2}\,\pl_{\bar{Y}_3}\,.
\end{align}
Without loss of generality, we can focus on $\delta^{(1)}_{E_1}\Phi_3$. The deformation of the gauge transformation can then be read off from the above formulas, giving
\begin{equation}
\delta^{(1)}_{E_1}\Phi(X,U)=-\Pi_\Phi\,T_{12}(X,U)\,,\label{gdef}
\end{equation}
where we introduced the projector $\Pi_\Phi$ to ensure the correct homogeneity degree and tangentiality conditions for the Fronsdal field $\Phi$ (c.f. \cite{Joung:2013nma}). 

\subsubsection*{Gauge algebra structure constants}

With the result \eqref{gdef} for $\delta^{(1)}$, as explained in \S \tcb{\ref{subsec::noerev}} the deformed structure constants of the gauge algebra can be extracted through \eqref{closure}
\begin{equation}
\delta^{(0)}_{[E_2}\delta^{(1)}_{E_1]}\Phi_3=U\cdot\pl_X\,[\![E_2,E_1]\!]_3^{(0)}\,.
\end{equation}
Referring the reader to \cite{Joung:2013doa} for further details, one obtains
\begin{multline}\label{str}
[\![E_1,E_2]\!]_3^{(0)}=\frac14\,\Pi_E\left(\pl_{Y_1}\pl_{ H_1}+\pl_{Y_2}\pl_{ H_2}\right)\\\left[\sum_{s_i}g_{s_1,s_2,s_3}\,e^{\bar{\mathfrak{D}}}\,Y_1^{s_1}Y_2^{s_2}Y_3^{s_3}\right]\,E_1(X_1,U_1)\,E_2(X_2,U_2)\Big|_{X_i=X,U_i=0}\,,
\end{multline}
where $\Pi_E$ ensures the correct homogeneity and tangentiality conditions for a gauge parameter. 

\subsection{Holographic higher-spin algebra structure constants: Test of the duality}
\label{subsec::hhasc}
The formula \eqref{str} obtained in the previous subsection gives the lowest order commutator for the gauge algebra of a putative higher-spin theory dual to the free scalar $O\left(N\right)$ model. Higher-spin symmetry is in this context gauged, and the corresponding global (or rigid) higher-spin symmetries can be obtained from evaluating the deformed bracket \eqref{str} on the gauge parameters $E = {\bar E}$ which satisfy the Killing equation 
\begin{equation}
U\cdot\pl_X\,\bar{E}(X,U)=0\,.\label{ke}
\end{equation}
In this section we show that these rigid structure constants indeed define a non-degenerate Lie algebra, coinciding with those of the Eastwood-Vasiliev higher-spin algebra \cite{Eastwood:2002su,Vasiliev:2004cm}. The Eastwood-Vasiliev algebra has in fact been shown to be unique in generic dimensions, through consideration of the Jacobi identity at the quartic order \cite{Boulanger:2013zza}. This result may therefore be considered as a test of the holographic duality, demonstrating that the holographically reconstructed theory is indeed the same theory one would obtain by solving the Noether procedure up to the quartic order.

\subsubsection*{Killing tensors}

We first review the solutions to the Killing equation \eqref{ke} in the framework of ambient space. Combined with the tangentiality and homogeneity conditions on the ambient space gauge parameter $E$,
\begin{equation}
    X \cdot \partial_U E\left(X,U\right) = 0, \qquad \left(X \cdot \partial_X - U \cdot \partial_U \right) E\left(X,U\right) = 0,
\end{equation}
it is straightforward to write down the Killing tensors ${\bar E}$, which read
\begin{equation}
\bar{E}(X,U)=\sum_{r=0}^\infty\,\frac{1}{(r!)^2}\,\bar{E}_{M_1N_1,\ldots,M_{r}N_r}\,X^{[M_1}U^{N_1]}\,\ldots\,X^{[M_r}U^{N_r]}.
\end{equation}
Combined with the tracelessness of the gauge parameter, one can also conclude that the Killing tensors are completely traceless 
\begin{equation}
    \partial_X \cdot \partial_U  {\bar E}\left(X,U\right)=0, \qquad \partial^2_X  {\bar E}\left(X,U\right)=0, \qquad \partial^2_U  {\bar E}\left(X,U\right)=0.
\end{equation}
The generators of a putative underlying higher-spin algebra are the duals of $\bar{E}_{M_1N_1,\ldots,M_{r}N_r}$, given by
\begin{equation}\label{gen}
 T^{M_1N_1,\ldots,M_{r}N_r}=X^{[M_1}U^{N_1]}\,\ldots\,X^{[M_r}U^{N_r]}+\ldots\,.
\end{equation}
 The $\ldots$ signify that the higher-spin generators, being contracted with traceless tensors, are defined as equivalence classes modulo traces: $X\cdot U$, $X^2$ and $U^2$. The $T^{M_1N_1,\ldots,M_{r}N_r}$ may thus be chosen to be traceless, with the symmetry of two row traceless $O\left(d,2\right)$ Young tableaux,
 \begin{equation}
     T^{M_1N_1,\ldots,M_{r}N_r}\quad \sim \quad 
	{\footnotesize\yng(7,7)}_{\,\circ}\ .
	\label{M yd}
 \end{equation}
 A generic killing tensor can therefore be parameterised by the following combination of null orthogonal auxiliary vectors $w_+$ and $w_-$,
 \begin{equation}
     W^{M N}  := w^{\left[M\right.}_+ w^{\left.N\right]}_{-}, \qquad w_\alpha \cdot w_\beta = 0, \quad \alpha, \beta = +, -.
 \end{equation}
To wit,
\begin{align}
\bar{E} &= \sum^\infty_{r=0} \frac1{2^{r}r!^2} T^{M_1N_1,\ldots,M_{r}N_r} W_{M_1 N_1} ...W_{M_r N_r} = \sum^\infty_{r=0} \frac1{2^{r}r!^2}[p_U m_X-m_U p_X]^{r}\,,\label{kt2}
\end{align}
where for ease of notation we have defined the following scalar contractions
\begin{align}
p_A&:=(w^+\cdot A)\,,& m_A := (w^-\cdot A)\,.
\end{align}
Computing the structure constants by evaluating \eqref{str} on Killing tensors \eqref{kt2} then boils down to an iterative application of the chain rule via
\begin{subequations}\label{chain}
\begin{align}
 H_i&=C_{jk}\,\pl_{p_{U_j}}\pl_{m_{U_k}}-C_{kj}\,\pl_{p_{U_k}}\pl_{m_{U_j}} + ... \,,\\
Y_1&=C_{12}\,\pl_{p_{U_1}}\pl_{m_{X_2}}-C_{21}\,\pl_{p_{X_2}}\pl_{m_{U_1}} + ...\,,\\
Y_2&=C_{12}\,\pl_{p_{U_2}}\pl_{m_{X_1}}-C_{21}\,\pl_{p_{X_1}}\pl_{m_{U_2}} + ...\,,\\
Y_3&=C_{31}\,\pl_{p_{U_3}}\pl_{m_{X_1}}-C_{13}\,\pl_{p_{X_1}}\pl_{m_{U_3}} + ...\,,
\end{align}
\end{subequations}
where we further defined $C_{ij}:= w^+_i\cdot w^-_j$. The $\ldots$ denote terms which may be neglected, as they are fixed by the symmetries of \eqref{M yd}. Employing the operator identities \eqref{chain}, in the following we extract the corresponding global symmetry structure constants.

\subsubsection*{Structure constants}
To extract the global symmetry structure constants induced by the holographic cubic action \eqref{full}, we evaluate the bracket \eqref{str} on the gauge parameters corresponding to Killing tensors \eqref{kt2}. By first considering the action of the projection operator $\Pi_E$ in \eqref{str}, a simplification is given by noting that the lowest derivative part of the commutator $[\![\bar{E}_1,\bar{E}_2]\!]_3^{(0)}$ has 
\begin{equation}
    \#\pl=s_1+s_2-s_3\,,
\end{equation}
corresponding to the correct degree of homogeneity in $X$. The remaining higher-derivative terms must be dressed by factors of $X^2$ to match the degree of homogeneity, and therefore just give rise to trace terms which can be set to zero in the quotient \eqref{gen}. The gauge-algebra commutator on the Killing tensors is thus the lowest derivative monomial
\begin{align}\label{strfin}
    f^{s_3}{}_{s_1,s_2}&:=[\![\bar{E}_{s_1},\bar{E}_{s_2}]\!]_{s_3}^{(0)}\\
    &=\frac{s_1!s_2!s_3!\,g_{s_1,s_2,s_3}}4\,\left(\pl_{Y_1}\pl_{ H_1}+\pl_{Y_2}\pl_{ H_2}\right)\nonumber\\
    &\times\sum_{n=0}^{s_3}\sum_{\alpha_1+\alpha_2\leq s_3-n}\tfrac{(-2)^n\lambda^{s_3-n}}{\ga_1!\ga_2!}\tfrac{ H_1^{s_3-\ga_2-n} H_2^{s_3-\ga_1-n}Y_1^{s_1-\ga_2-n}Y_2^{s_2-\ga_1-n} G^{s_3-\ga_1-\ga_2}}{(s_1-\ga_2-n)! (s_2-\ga_1-n)!  (s_3-\ga_1-\ga_2-n)! (2 n-s_3+\ga_1+\ga_2)!}\nonumber\\
    &\times\,E_1(X_1,U_1)\,E_2(X_2,U_2)\Big|_{X_i=X,U_i=0}\,.\nonumber
\end{align}
Evaluating the derivatives in above to obtain its explicit form is straightforward, but lengthy in general. One can proceed by expanding every term, performing all differentiations and re-summing. The final result can be expressed in terms of four basic traces
\begin{subequations}\label{Ms}
\begin{align}
M_{ij}&=\text{Tr}(W_iW_j)\,,\\ M_{123}&=\text{Tr}(W_1W_2W_3)\,,
\end{align}
\end{subequations}
which parameterise the most general decomposition of the trace of the triple tensor product of three two-row traceless window Young tableaux
\begin{multline}
   \text{Tr}\left( M_1^{s_1-1}\otimes M_2^{s_2-1}\otimes M_3^{s_3-1}\right)\\\sim \quad \sum c_k^{(s_1,s_2,s_3)}\, M_{12}^{\tfrac{s_1+s_2-s_3-1-k}2}M_{23}^{\tfrac{s_2+s_3-s_1-1-k}2}M_{31}^{\tfrac{s_3+s_1-s_2-1-k}2}M_{123}^k\,.
\end{multline}

\subsubsection*{Invariant form and cyclic structure constants}
In order to compare our result obtained via holography with the known structure constants for higher-spin lie algebras (see \S \tcb{\ref{appendix}}), we require the cyclic structure constants $f_{s_1,s_2,s_3}$. These can be obtained with the knowledge of the corresponding invariant form $\kappa_{s,s^\prime}$,
\begin{equation}
f_{s_1,s_2,s_3}\equiv \sum_sk_{s_1,s}f^{s}{}_{s_2,s_3}=\left\langle\bar{E}_{s_1}|[\bar{E}_{s_2},\bar{E}_{s_3}]\right\rangle.\label{cyc}
\end{equation}
Without loss of generality, the invariant form can be chosen to take the diagonal form
\begin{equation}
\kappa_{s,s^\prime}=\left\langle\bar{E}_{s}|\bar{E_{s^\prime}}\right\rangle\equiv\delta_{s,s^\prime}\,b_s\,\frac{(\pl_{U_1}\cdot\pl_{U_2})^{s-1}}{(s-1)!}\frac{(\pl_{X_1}\cdot\pl_{X_2})^{s-1}}{(s-1)!}\,\bar{E}_1(X_1,U_1)\,\bar{E}_2(X_2,U_2)\Big|_{X_i=0,U_i=0}\,,\label{bif}
\end{equation}
for some constants $b_s$, which can be fixed uniquely up to an overall coefficient by enforcing cyclicity of $f_{s_1,s_2,s_3}$. 

For example, they can be determined simply by considering the structure constants induced by the minimal gravitational coupling, which entails solving the equation
\begin{equation}
f_{2ss}\equiv\left\langle\bar{E}_{2}|[\bar{E}_{s},\bar{E}_{s}]\right\rangle=\left\langle\bar{E}_{s}|[\bar{E}_{2},\bar{E}_{s}]\right\rangle\equiv f_{s2s}\,
\end{equation}
for the coefficients $b_s$ contained in the definition \eqref{cyc}. In this way we obtain
\begin{equation}\label{bcoeff}
b_s=\frac{\pi ^{\frac{d-1}{2}}(-1)^{s-1} 2^{d-s+5} \Gamma \left(\frac{d+1}{2}\right)}{\Gamma \left(\frac{d}{2}-1\right) \Gamma \left(\frac{d}{2}+s-2\right)}\,,
\end{equation}
where the overall constant has been fixed by normalising the $1$-$1$-$1$ structure constants to the identity.
This leads to the diagonal bi-linear form
\begin{equation}
\kappa_{s,s} = \text{Tr}(T_s\star T_s)=\frac{\pi ^{\frac{d-1}{2}}\, 2^{d-4 s+8}\,s\, \Gamma \left(\frac{d+1}{2}\right)}{\Gamma \left(\frac{d}{2}-1\right) \Gamma \left(\frac{d}{2}+s-2\right)}\,\frac{1}{(s-1)!^2}\,M_{12}^{s-1}\,.
\end{equation}
The corresponding cyclic structure constants are then 
\begin{multline}\label{strconst}
f^{(s_1,s_2,s_3)}= g_{s_1,s_2,s_3}\sum_{l=0[1]}^{\text{Min}(s_1,s_2,s_3)-1}\bar{f}^{((s_1+s_2-s_3-l-1)/2,(s_2+s_3-s_1-l-1)/2,(s_3+s_1-s_2-l-1)/2,l)}\\\times\,M_{12}^{(s_1+s_2-s_3-l-1)/2}M_{23}^{(s_2+s_3-s_1-l-1)/2}M_{31}^{(s_3+s_1-s_2-l-1)/2}M_{123}^l\,,
\end{multline}
where the sum over $l$ ranges over the odd integers if $s_1+s_2+s_3$ is even, and over even integers if $s_1+s_2+s_3$ is odd. The coefficients ${\bar f}^{(a,b,c)}$ are defined by

{\footnotesize
\begin{align}\label{strAdSd}
{\bar f}^{(k_1,k_2,k_3,l)}&=\sum_{n=0}^\infty\sum_{m=0}^\infty \mathcal{P}_{m,n}^{(k_i,l)} \pi ^{\frac{d}{2}} (-1)^m \Gamma \left(\tfrac{d+1}{2}\right) (k_1+k_2+l+1) (k_1+k_3+l+1) (k_2+k_3+l+1)\\
&\hspace{-40pt}\frac{2^{-3 k_1-3 k_2-3 k_3-4 l-2 m-6 n+5} \Gamma (m+2 n+2) \Gamma \left(\frac{d}{2}+m+2 n-\frac{1}{2}\right) \Gamma (d+2 k_1+2 k_2+2 k_3+3 l)}{\Gamma \left(\frac{d}{2}-1\right) \Gamma \left(\frac{d+1}{2}+m+3 n\right)\Gamma \left(\frac{d}{2}+k_1+k_2+l-1\right) \Gamma \left(\frac{d}{2}+k_1+k_3+l-1\right) \Gamma \left(\frac{d}{2}+k_2+k_3+l-1\right)}\,,\nonumber
\end{align}}
\noindent
\!\!with

{\footnotesize
\begin{align}
\mathcal{P}_{m,n}^{(k_i,l)}&=\frac{1}{\Gamma (-k_1-k_2-l+m+2 n+1) \Gamma (-k_1-k_3-l+m+2 n+1) \Gamma (-k_2-k_3-l+m+2 n+1)}\nonumber\\
&\times\frac1{\Gamma (k_1+k_2+k_3+l-m-2 n+1) \Gamma (-k_1-k_2-k_3-l+m+3 n+1)}\,\nonumber\\
&\times\frac1{\Gamma (2 k_1+2 k_2+2 k_3+3 l-2 m-6 n+1)}\,.
\end{align}}
\noindent
\!\!For simplicity the sum over $n$ and $m$ is extended up to infinity owing to the poles of the $\Gamma$-functions in the denominators of $\mathcal{P}_{m,n}^{(k_i,l)}$. Therefore only finitely many terms contribute to the above sums for a given triplet of spin.\footnote{The ranges can be straightforwardly recovered by solving the  inequalities:
\begin{subequations}
\begin{align}
m+2n&\geq \text{Max}(s_i)-1\,,\\
m+2n&\leq \frac{s_1+s_2+s_3-3}{2}-l\,,\\
m+3n&\leq \frac{s_1+s_2+s_3-3}{2}\,,\\
m+3n&\geq \frac{s_1+s_2+s_3-3}{2}-l\,.
\end{align}
\end{subequations}}

Although the result \eqref{strconst} is lengthy, the same complicated expression is obtained from expanding the known generating functions for the structure constants of the $hs\left(so\left(d-1,2\right)\right)$ higher-spin lie algebra (see \S \tcb{\ref{appendix}}). In particular, since the latter algebra is unique in generic dimensions,\footnote{In AdS$_3$ and AdS$_5$ there are one-parameter families of higher-spin algbras, and in these dimensions the structure constants \eqref{strconst} coincide with the known expressions for parameters which correspond to the symmetries of the free scalar theory on the boundary. I.e. the test is also passed in these cases.} this verifies that the cubic couplings \eqref{full} obtained holographically give the same deformations of the gauge symmetries as those which would be obtained from the Noether procedure at quartic order independently of holography. This extends to general dimensions the tree-level three-point function test \cite{Giombi:2009wh} of higher-spin holography by Giombi and Yin in AdS$_4$.

\section{Higher-spin cubic couplings in 4d flat space}\label{Metsaev}
For the remainder of this article we turn to higher-spins in flat space. Higher-spin cubic couplings which solve the Noether procedure up to the second non-trivial order (quartic) were first studied in the early 90s by Metsaev \cite{Metsaev:1991nb,Metsaev:1991mt}, using light-cone methods.\footnote{This postdated the original cubic classification of \cite{Bengtsson:1983pd,Bengtsson:1986kh}, solving the Noether procedure at the first non-trivial order.} In the light-cone gauge, the Noether procedure reduces to requiring the closure of the Poincar\'e generators deformed by cubic interactions. In this way the light-cone Lagrangian can be read off from the non-linear deformation of the light-cone Hamiltonian. The quartic order analysis of \cite{Metsaev:1991nb,Metsaev:1991mt} remarkably led to the complete fixing of the flat space cubic action in four dimensions. In this section we analyse this cubic theory and explore a putative underlying higher-spin algebra in four-dimensional flat space, extending the discussion carried out in the previous section.

\subsection{Light-cone gauge}
\label{lcdic}
We first review the gauge fixing of higher-spin fields in flat space to  light-cone gauge. It is convenient to work with the light-cone coordinates 
\begin{align}
x^\pm&\equiv \frac{x^0\pm x^3}{\sqrt{2}}\,,& z&=\frac{x^1+i x^2}{\sqrt{2}}\,,& \bar{z}&=\frac{x^1-i x^2}{\sqrt{2}}\,,
\end{align}
 with diagonal metric $$g_{\mu\nu}=(-1,+1,+1,+1)\,.$$
The corresponding space-time derivatives are given by
\begin{align}
\pl^\pm x^\mp&=g^{\pm \mp}\,,& \pl\, \bar{z}&=g^{z\bar{z}}\,,& \bar{\pl}\,
z&=g^{\bar{z}z}\,.
\end{align}
By introducing auxiliary variables $u^\mu=(u^+,u^-,u,\bar{u})$, the usual Fierz system which packages higher-spin fields\footnote{Recall the generating function $\varphi(x,u)$ encodes spin-$s$ fields \begin{equation}
   \varphi(x,u) = \sum_{s} \frac{1}{s!} \varphi_{\mu_1...\mu_s} u^{\mu_1} ...u^{\mu_s}.
\end{equation}}
\begin{align}
    \Box \varphi(x,u)=0\,,\qquad \partial \cdot \partial_u \varphi(x,u)=0\,,\qquad
\partial^2_u\varphi(x,u)=0\,,
\end{align}
becomes
\begin{subequations}\label{lcfierz}
\begin{align}
(-2\pl^+\pl^-+2\pl\bar{\pl})\varphi(x,u)&=0\,,\\
(-\pl^+\pl_u^--\pl^-\pl_u^++\pl\bar{\pl}_u+\bar{\pl}\pl_u)\varphi(x,u)&=0\,,\\
(-2\pl_u^+\pl_u^-+2\pl_u\bar{\pl}_u)\varphi(x,u)&=0\,.
\end{align}
\end{subequations}
The leftover gauge symmetry
\begin{equation}
\delta_\xi \varphi(x,u)=(-u^+\pl^--u^-\pl^++u\bar{\pl}+\bar{u}\pl)\,\xi\,,
\end{equation}
can be completely fixed by requiring $\pl_u^+\,\varphi(x,u)=0$, for which the system \eqref{lcfierz} becomes
\begin{subequations}
\begin{align}
(-2\pl^+\pl^-+2\pl\bar{\pl})\varphi(x,u)&=0\,,\\
\left(\pl^+\pl_u^--\pl\,\bar{\pl}_u-\bar{\pl}\,\pl_u\right)\varphi(x,u) & = 0\,,\label{dcond}\\
\pl_u\bar{\pl}_u\,\varphi(x,u)&=0\,.\label{trcond}
\end{align}
\end{subequations}
By solving the divergence condition \eqref{dcond}
\begin{equation}
\pl_u^-\varphi(x,u)\rightarrow \frac1{\pl^+}(\pl\,\bar{\pl}_u+\bar{\pl}\,\pl_u)\varphi(x,u),
\end{equation}
together with the traceless constraint \eqref{trcond} this enables the fields to be expressed in terms of the two physical helicities in four-dimensions
\begin{equation}
    \varphi(x,u)=\sum_s{\varphi}_{-s}(x)\,u^s+\varphi_{+s}(x)\,\bar{u}^s,\label{philc}
\end{equation}
which are encoded by a pair of complex conjugate scalar fields ${\varphi}_{-s}(x)\equiv \varphi_{\bar{z}(s)}(x)$ and $\varphi_s(x)\equiv \varphi_{{z}(s)}(x)$.

\subsection{Cubic vertices in light-cone gauge}\label{gaugevertex}
We now apply the dictionary \S \tcb{\ref{lcdic}} for expressing physical quantities in light cone gauge to the most general cubic vertex. At first non-trivial order in the Noether procedure (which leaves the relative coefficients unfixed), gauge invariant parity preserving cubic vertices in flat space have been classified in a manifestly covariant form in \cite{Manvelyan:2010je,Sagnotti:2010jt,Fotopoulos:2010ay}. Their general structure in TT (traceless and transverse, c.f. \S \tcb{\ref{subsubsec::TT}}) gauge is given in generating function notation by
\begin{equation}
    \mathcal{V}_3=f(Y_1,Y_2,Y_3,G)\,\varphi_1(x_1,u_1)\varphi_2(x_2,u_2)\varphi_3(x_3,u_3)\Big|_{u_i=0,x_i=x}\,,\label{flatC}
\end{equation}
with $Y_i$ and $G$ defined as
\begin{align}
Y_i&=\pl_{u_i}\cdot\pl_{x_{i+1}}\,,& G&=\pl_{u_1}\cdot\pl_{x_2}\,\pl_{u_2}\cdot\pl_{u_3}+\text{cyclic}\,.\label{flatbb}
\end{align}
These are the flat space analogues of the AdS building blocks \eqref{adsbb} and \eqref{gads}. 
The light cone gauge-fixing can be carried out directly at the level of \eqref{flatbb}, and is achieved for any term in \eqref{flatC} in combination with \eqref{philc} simply by replacing
\begin{align}
Y_i&\rightarrow -(\pl_{x_i}^+)^{-1}\left[\bar{P}\pl_{u_i}+P\bar{\pl}_{u_i}\right]\,,\\
G&\rightarrow\bar{\pl}_{u_1}\bar{\pl}_{u_2}{\pl}_{u_3}\,\left(\frac{\pl_{x_3}^+}{\pl_{x_1}^+\pl_{x_2}^+}\,P\right)+{\pl}_{u_1}{\pl}_{u_2}\bar{\pl}_{u_3}\,\left(\frac{\pl_{x_3}^+}{\pl_{x_1}^+\pl_{x_2}^+}\,\bar{P}\right)+\text{cyclic}\,,
\end{align}
where we have introduced (anti-)holomorphic light-cone momenta $P ({\bar P})$:
\begin{align}\label{pbp}
P&=\frac13\left[\pl_{x_1}\,(\pl^+_{x_2}-\pl^+_{x_3})+\pl_{x_2}\,(\pl^+_{x_3}-\pl^+_{x_1})+\pl_{x_3}\,(\pl^+_{x_1}-\pl^+_{x_2})\right]\,,\\
\bar{P}&=\frac13\left[\bar{\pl}_{x_1}\,(\pl^+_{x_2}-\pl^+_{x_3})+\bar{\pl}_{x_2}\,(\pl^+_{x_3}-\pl^+_{x_1})+\bar{\pl}_{x_3}\,(\pl^+_{x_1}-\pl^+_{x_2})\right]\,.
\end{align}

\subsection*{Parity violating vertices in light-cone gauge}
In four-dimensions the epsilon tensor can be used to construct parity violating gauge-invariant vertices. The basic parity violating structures are given (up to integration by parts) by
\begin{equation}
\mathcal{E}_i=\epsilon_{\mu\nu\rho\sigma}
\pl_{u_j}^\mu \pl_{u_k}^\nu \pl_{x_i}^\rho \pl_{x_j}^\sigma\,,
\end{equation}
with $i$, $j$ and $k$ cyclically ordered. 

The most general parity violating vertex will then also depend on the above additional structures, multiplied by an arbitrary parity even structure built from \eqref{flatbb} introduced in the previous section. 

The light-cone gauge fixing in this case is obtained through the replacement
\begin{equation}
\mathcal{E}_i\rightarrow (\pl_{x_j}^+)^{-1}(\pl_{x_k}^+)^{-1}\,\left[\bar{\pl}_{u_j}\bar{\pl}_{u_k}\,P^2-\pl_{u_j}\pl_{u_k}\,\bar{P}^2\right]\,.
\end{equation}
Parity violating deformations of Metsaev's solution \cite{Metsaev:1991nb,Metsaev:1991mt} are expected to have fixed overall coefficients, with the only dependence being on one parity violating parameter analogous to the $\theta$ parameter in the AdS$_4$ theory (see e.g. \cite{Sezgin:2002ru}). 

In this paper we shall only consider parity preserving cubic couplings.

\subsection{Metsaev's cubic action}
\label{subsec::mca}
In the previous section, we considered the light-cone gauge fixing of cubic structures \eqref{flatC} obtained by solving the Noether consistency conditions with manifest covariance at first non-trivial order. At this order, the relative coefficients between the independent cubic structures are unfixed. On the other hand, purely within the light-cone framework, Metsaev fixed the cubic action completely almost 25 years ago \cite{Metsaev:1991nb,Metsaev:1991mt}. In terms of the light-cone momenta $P$ and $\bar{P}$ defined by \eqref{pbp}, it takes the rather simple form
\begin{multline} \label{mcub}
\mathcal{V}_3=\sum_{|s_1|,|s_2|,|s_3|}\Big[\frac{(il)^{s_1+s_2+s_3}}{\Gamma(s_1+s_2+s_3)}\,(\pl_{x_1}^+)^{-s_1}(\pl_{x_2}^+)^{-s_2}(\pl_{x_3}^+)^{-s_3}\,\bar{P}^{s_1+s_2+s_3}\varphi_{+s_1}\varphi_{+s_2}\varphi_{+s_3}\\+\frac{(-il)^{-s_1-s_2-s_3}}{\Gamma(-s_1-s_2-s_3)}\,(\pl_{x_1}^+)^{s_1}(\pl_{x_2}^+)^{s_2}(\pl_{x_3}^+)^{s_3}\,{P}^{-s_1-s_2-s_3}\varphi_{-s_1}\varphi_{-s_2}\varphi_{-s_3}\Big]\,,
\end{multline}
with the sum running over all integer helicities $\pm s$.\footnote{It is interesting to note that the $\Gamma$-function coupling constant ensures strict light-cone locality: Whenever $s_1+s_2+s_3<1$ or $-s_1-s_2-s_3<1$, the coefficient of a putative ``non-local'' structure (i.e. with $1/\partial^{\pm}_x$) vanishes identically.}. Above we have introduced the coupling constant $l$ to dress the higher-derivative interactions. We discuss a few notable properties of the action in the following. 

For given triplet of spins $(s_1,s_2,s_3)$, in \eqref{mcub} there are as many structures as the number of positive combinations $\pm s_1\pm s_2\pm s_3$, giving rise to three distinct cases:
\begin{enumerate}
\item  \underline{$s_1=s_2=s_3=s$}

There are two possible such combinations, one with $3s$ derivatives and another with $s$ derivatives. The latter for $s=1,2$ reproduces the standard Yang-Mills self interaction and the Einstein-Hilbert minimal coupling, respectively.
\item \underline{$s_1=s_2=s$ and $s_3\neq s$} 

Here the first unexpected feature emerges. In this case there are \emph{three} different couplings, with $2s+s_3$, $|2s-s_3|$ and $s_3$ derivatives. This is in contrast to the expected number (two) of couplings, obtained from a covariant classification \cite{Manvelyan:2010jr,Sagnotti:2010jt}. For example, here in the gravitational case $s-s-2$ we have the expected couplings with $2s+2$ and $2s-2$ derivatives, but also a third with $2$ derivatives. The latter may be considered as the gravitational minimal coupling and was referred as exotic in \cite{Metsaev:2005ar}. In the following sections, by examining corresponding deformations of the gauge symmetries, we will argue that this two-derivative coupling may indeed be given the interpretation of minimal coupling.
\item  \underline{$s_1\ne s_2 \ne s_3$} 

This case gives even more surprises, where the number of independent couplings grows up to four, with $s_1+s_2+s_3$, $|s_1+s_2-s_3|$, $|s_2+s_3-s_1|$ and $|s_3+s_1-s_2|$ derivatives. This is to be compared with the covariant classification of cubic couplings, where only two couplings could be identified with $s_1+s_2+s_3$ and $s_1+s_2+s_3-2s_{\text{min}}$ derivatives. The additional couplings were also referred as exotic in \cite{Metsaev:2005ar}. 
\end{enumerate}

In the following sections we study the ``additional'' vertices highlighted in 2. and 3. above and their possible implications.

\subsection{Covariantising Metsaev's vertices}
\label{subsec::covmetv}
In this section we revisit the covariant classification of cubic vertices (c.f. \S \tcb{\ref{gaugevertex}}), with the aim of accommodating the additional vertices reviewed in the previous section, which were discovered in the light-cone gauge. The latter were previously unaccounted for in the original covariant classification \cite{Boulanger:2008tg,Manvelyan:2010jr,Sagnotti:2010jt,Fotopoulos:2010ay}. Furthermore, in this way we may apply covariant methods such as those in \S \tcb{\ref{subsec:::defgaugesymm}} and \S \tcb{\ref{subsec::hhasc}} for investigating a putative flat space higher-spin algebra.  

In flat space,\footnote{c.f. \S \tcb{\ref{subsubsec::TT}} for the AdS analogue. In particular, they differ by a factor $e^{\mathfrak{D}}$ where $\mathfrak{D}$ is the differential operator \eqref{mfd} which generates corrections to the flat space result from non-zero curvature.} the most general gauge invariant cubic structure can be parameterised by the building blocks
\begin{equation}\label{genk}
f_{s_1,s_2,s_3}^{(k)} := Y_1^{s_1}Y_2^{s_2}Y_3^{s_3}\left(\frac{G}{Y_1\,Y_2\,Y_3}\right)^k\,,
\end{equation}
for fixed external spins $(s_1,s_2,s_3)$. Note that the structures \eqref{genk} are polynomial in the oscillators only if $k\leq \text{min}(s_1,s_2,s_3)$.
If $k >\text{min}(s_1,s_2,s_3)$, the covariant expression is still formally gauge invariant and the light-cone gauge-fixing described in \S \tcb{\ref{gaugevertex}} can be formally applied.

The key observation, which was not considered in the original covariant classification, is that although the structures \eqref{genk} for $k >\text{min}(s_1,s_2,s_3)$ are formally non-polynomial in the oscillators $\pl_u$ and ${\bar \pl_u}$, all non-polynomial dependence cancels out after gauge fixing to the light-cone.\footnote{This essentially due to the factorised form of the light cone traceless condition
\begin{equation}
\pl_u\bar{\pl}_u\,\varphi=0\,,
\end{equation}
whose solution are either holomorphic or anti-holomorphic in the variable $u$.} More explicitly, employing the dictionary given in \S \tcb{\ref{gaugevertex}}, on the light-cone the structures \eqref{genk} for general $k$ read\footnote{For simplicity we display the structures proportional to $\bar{\pl}_{u_1}^{s_1}\bar{\pl}_{u_2}^{s_2}\bar{\pl}_{u_3}^{s_3}$ and $\bar{\pl}_{u_1}^{s_1}\bar{\pl}_{u_2}^{s_2}\pl_{u_3}^{s_3}$, the remaining two can be obtained analogously.} 
\begin{align}
&(f_{s_1,s_2,s_3}^{(0)})_{---}\rightarrow (\pl_{x_1}^+)^{-s_1}(\pl_{x_2}^+)^{-s_2}(\pl_{x_3}^+)^{-s_3}\,{P}^{s_1+s_2+s_3}\,\bar{\pl}_{u_1}^{s_1}\bar{\pl}_{u_2}^{s_2}\bar{\pl}_{u_3}^{s_3}\,,\\
&(f_{s_1,s_2,s_3}^{(k)})_{--+} \nonumber \\ & \hspace*{1cm} \rightarrow (-1)^{s_1+s_2+s_3-k}\,(\pl_{x_1}^+)^{-s_1}(\pl_{x_2}^+)^{-s_2}(\pl_{x_3}^+)^{2k-s_3}P^{s_1+s_2-k}\bar{P}^{s_3-k}\,\bar{\pl}_{u_1}^{s_1}\bar{\pl}_{u_2}^{s_2}\pl_{u_3}^{s_3}\,,
\end{align}
which are indeed polynomial in $\pl_u$ and ${\bar \pl_u}$. In the following we thus relax the constraint $k\leq \text{min}(s_1,s_2,s_3)$, and explore the structures in \eqref{genk} which may accommodate consistent cubic interactions.

We first note that the structure $f_{s_1,s_2,s_3}^{(k)}$ is holomorphic only for $k=s_3$ or $k=s_1+s_2$, regardless if $s_3=s_{\text{min}}$ or not. The non-holomorphic terms can either be removed by a local field redefinition when they are proportional to $P\bar{P}$, or by a non-local but admissible redefinition (\`a la \S \tcb{\ref{subsec::lcl}}) when they are of the form $P^n/\bar{P}^m$ or $\bar{P}^n/{P}^m$ with both $n\neq0$ and $m\neq 0$. The remaining terms which cannot be removed are non-local, but can be avoided by placing restrictions on the value of $k$ as we discuss below.

\subsubsection*{Locality}
Although in the above we relaxed the constraint $k\leq \text{min}(s_1,s_2,s_3)$, locality places restrictions on the range of $k$, which we consider here. 

The only non-local terms which cannot be removed by admissible redefinitions, and which would give rise to a singular S-matrix, are those of the type $\frac1{P^n\bar{P}^m}$ for $n\geq0$ and $m\geq0$ excluding the constant (for the details, see \S \tcb{\ref{subsec::lcl}}). This can be avoided requiring 
\begin{align}
k\leq\text{min}(s_1+s_2,s_3)\,,
\end{align}
which upon cyclising the indices can be rewritten as
\begin{equation}
    s_1+s_2+s_3-2k\geq0. \label{lok}
\end{equation}

With the locality condition \eqref{lok} satisfied and allowing field redefinitions of the type described in \S \tcb{\ref{subsec::lcl}}, we are left with only holomorphic or anti-holomorphic terms which cannot be removed by a redefinition. Discarding couplings which do not give rise to (anti-) holomorphic structures gives the following list of covariant couplings for fixed spin:
\begin{align}
&f_{s_1,s_2,s_3}^{(0)}\,,& &f_{s_1,s_2,s_3}^{(k)}\,,& s_1+s_2+s_3-2k\geq0\,.
\end{align}
with a number of different local couplings equal to the number of unequal spins plus one. The two couplings which fall into the original covariant classification of \cite{Boulanger:2008tg,Manvelyan:2010jr,Sagnotti:2010jt,Fotopoulos:2010ay} are given by
\begin{align}
f_{s_1,s_2,s_3}^{(0)}\, \qquad \text{and} \qquad f_{s_1,s_2,s_3}^{(s_{\text{min}})}\,.
\end{align}

\subsubsection*{Covariant cubic action}
Combining the above light-cone $\rightarrow$ covariant dictionary, we obtain the following (formal)\footnote{We emphasise that this re-writing of Metsaev's vertices is strictly formal, and serves primarily as an auxiliary step to extract the higher-spin structure constants. On the other hand, it is possible to enlarge the functional space of polynomials $\phi_{\mu(s)}(x)u^{\mu(s)}$ and allow $1/Y$ poles, in spite of the lack of tensorial interpretation. Within such a non-tensorial functional space, \eqref{MetCov} represents the covariantisation of Metsaev vertices.} covariant rewriting of Metsaev's vertices up to the class of re-definitions given in \S \tcb{\ref{subsec::lcl}}:\footnote{The kinetic term is normalised with $\frac{1}{2^s s!}$ for convenience.}
\begin{align}
\mathcal{V}_3&=\mathcal{M}(Y_1,Y_2,Y_3,G)\,\varphi_1\,\varphi_2\,\varphi_3\,,\\
\mathcal{M}&=\sum_{s_i=0}^\infty\left[\sum_{k=\left\{0,j\right\}}\frac{(il)^{s_1+s_2+s_3-2k}}{\sqrt{s_1!s_2!s_3!}\,\Gamma(s_1+s_2+s_3-2k)}\,Y_1^{s_1}Y_2^{s_2}Y_3^{s_3}\left(\frac{G}{Y_1\,Y_2\,Y_3}\right)^k\right]\,,\label{MetCov}
\end{align}
with $j$ spanning the three values satisfying
\begin{equation}
    s_1+s_2+s_3-2j = |s_p+s_q-s_r|, \qquad p \ne q \ne r.
\end{equation}
We now comment on the properties of \eqref{MetCov} in contrast to the original covariant classification, in which additional vertices highlighted in \S\tcb{\ref{subsec::mca}} did not appear. As may be anticipated, the covariant form of these additional vertices  contain poles in $Y_i$. For example, the two-derivative gravitational coupling is given explicitly by
\begin{equation}
\frac{G^s}{Y_3^{s-2}}=G^2\left(H_3+\frac{H_1Y_1+H_2Y_2}{Y_3}\right)^{s-2}\,,\label{gocv}
\end{equation}
with coupling constant fixed in \eqref{MetCov}. Notice that all spin-$s$ two-derivative gravitational coupling constants are indeed \emph{equal}:
\begin{equation}
    \frac{(il)^{2s+2-2k}}{\Gamma(2s+2-2k)}\Big|_{k=s}=(il)^{2}\,,
\end{equation}
and spin-independent in accordance with the equivalence principle \cite{Weinberg:1964ew}. We emphasise that that all apparent singularities of the above non-polynomial solutions to the Noether procedure disappear upon gauge fixing to the light cone in $4d$. This suggests that the non-local singular covariant form \eqref{gocv} might just be an artifact of choosing not to introduce auxiliary fields to solve for gauge consistency. Indeed, the reason why the above vertices were overlooked in the original treatment is that they do not admit a standard tensorial form. Taking into account these caveats, let us stress that we only use this rewriting as a formal trick to extract the structure constants using covariant methods (such as those in \S\tcb{\ref{AdS}}). As we demonstrated above, a non-singular formulation of the exotic vertices is currently only available in a Lorentz non-covariant frame (i.e. in light-cone gauge).

\subsection{Light-cone locality}
\label{subsec::lcl}
In this section we detail the class of field-redefinitions used to obtain the formal covariant re-writing \eqref{MetCov} of the light-cone cubic action \eqref{mcub}.

Using the identity:
\begin{equation}
2P\bar{P}=-\Box_1\,\pl_{x_2}^+\pl_{x_3}^+-\Box_2\,\pl_{x_3}^+\pl_{x_1}^+-\Box_3\,\pl_{x_1}^+\pl_{x_2}^++2\,(\pl_{x_1}^-+\pl_{x_2}^-+\pl_{x_3}^-)\pl_{x_1}^+\pl_{x_2}^+\pl_{x_3}^+\,,
\end{equation}
we can see that the combination $P\bar{P}$ (up to total derivatives) is proportional to the equations of motion, and for this reason can be removed by a field redefinition. In particular, going on-shell in light cone gauge is equivalent to setting
\begin{equation}
P\bar{P}\approx 0\,.\label{pp1}
\end{equation}
The above equation factorises in four dimensions,\footnote{In higher dimensions $d$ one works with a $so(d-2)$ vector $P^I$ and factorisation would break $so(d-2)$ covariance.} and so it can be solved in two possible ways:\footnote{On-shell this recovers in disguise the well-known holomorphic structures usually found in the spinor-helicity formalism (see e.g. \cite{Benincasa:2007xk} and also \S \tcb{\ref{Sec:spinohelicity}}.).}
\begin{subequations} \label{fact}
\begin{align}
    \text{1.} \quad P\neq 0\rightarrow \bar{P}=\frac1{P}\,P\bar{P}\approx 0\,,\\
    \text{2.} \quad \bar{P}\neq 0\rightarrow P=\frac1{\bar{P}}\,P\bar{P}\approx 0\,,
\end{align}
\end{subequations}
Notice that for both $P=0$ and $\bar{P}=0$ no non-singular and non-trivial solution can be written down.

Interestingly, the above observation implies that formally one can relax the light-cone locality condition which requires the vertex to be a polynomial in $P$ and $\bar{P}$ (at least for any fixed triple of spins), since a non-singular branch in \eqref{fact} can always be chosen if one of them is zero on-shell. Using this observation, we show that in this case there exists an enlarged class of field re-definitions which leave the S-matrix invariant. These are not globally defined as they are singular for generic on-shell configurations, however they are non-singular on one branch \eqref{fact} of the on-shell surface at a time. In \S \tcb{\ref{subsec::covmetv}} this enlarged class of re-definitions enabled a formal covariantisation of the exotic light-cone vertices in a particular field frame.

In order to discuss these issues, we recall the important requirement that the S-matrix of a theory should be finite for generic on-shell configurations.
Since combinations of the type $P\bar{P}$ vanish on-shell usually in the light cone gauge one has only holomorphic or anti-holomorphic local vertices. For fixed external spins, the S-matrix in this case is thus polynomial in the light-cone momenta $P$ and $\bar{P}$. 

In four-dimensions, the factorisation property \eqref{fact}, which gives a factorised on-shell surface:
\begin{equation}
    \mathcal{S}=\{P=0\}\oplus\{\bar{P}=0\}\,,
\end{equation}
permits a wider class of vertices: Consider a vertex of the type\footnote{The discussion proceeds in the same way for vertices of the form \begin{equation}
    \frac{{P}^n}{{\bar P}^m}.
\end{equation}}
\begin{equation}
    \frac{{\bar P}^n}{P^m}, \qquad \text{with} \quad m > 0,\label{newv}
\end{equation}
which are non-polynomial in one of the light-cone momenta. There are two distinct cases to consider: $n = 0$ and $n > 0$. For $n = 0$, while such vertices yield a non-singular S-matrix on the branch $\{{\bar P}=0\}$, they are singular on the branch $\{ P=0\}$ and are thus excluded. For $n > 0$, however, there is a crucial difference: Although this type of vertex is also singular on $\{ P=0\}$, on the branch $\{{\bar P}=0\}$ they are proportional to the equations of motion:
\begin{equation}
    \frac{{\bar P}^n}{P^m} =  \frac{{\bar P}^n}{P^m} \cdot \frac{P}{P} = {\bar P}P \cdot \frac{{\bar P}^{n-1}}{P^{m+1}},
\end{equation}
and can be removed by a field redefinition. This redefinition is not globally defined, but however finite on one branch of the on-shell surface. Motivated by this observation, it seems reasonable to allow non-local vertices of the type \eqref{newv} for $n>0$, which arise from choices of the field-variables which may not be globally defined on the full on-shell surface but still well-defined on either of the branches $\{P=0\}$ and $\{\bar{P}=0\}$. 

Let us stress that the above functional class, although enlarged compared to the generic case, allowing such \emph{singular} redefinitions do not remove on-shell non-trivial local vertices. For example a vertex proportional to $P^n$ which lives on the $\{{\bar P}=0\}$ branch of the stationary surface cannot be removed by a redefinition of the type ${\bar{P}}/{P}$ on the same branch. In particular, multiplying or dividing by $\bar{P}$ is not allowed in this functional space for field configurations $\bar{P}\approx 0$. 

\subsection{Deformations of Gauge Symmetries from Metsaev vertices}
With a covariant form \eqref{MetCov} of the couplings \cite{Metsaev:1991mt,Metsaev:1991nb} established, we can extract the corresponding deformations of the gauge transformations and their commutators by employing covariant formulas, as in the AdS case \S \tcb{\ref{AdS}}. We further extract the structure constants of a putative higher-spin algebra and discuss the result. 

\subsubsection*{Gravitational coupling of higher-spins in flat space}
We first consider the gravitational coupling of spin-$s$ gauge fields, in particular in the view of the two-derivative $s$-$s$-$2$ coupling highlighted in \S \tcb{\ref{subsec::mca}}. We extract the structure constants of the semi-simple (higher-spin Lorentz) subalgebra of the putative higher-spin algebra, and argue that the latter two-derivative coupling can be interpreted as a minimal coupling of higher-spin gauge fields to gravity.

To this end it is straightforward to apply the same techniques employed for the AdS case in \S \tcb{\ref{subsec:::defgaugesymm}}. The deformed gauge bracket is 
\begin{equation}
    [\![\xi_1,\xi_2]\!]_3^{(0)}=\frac14\,\Pi_\xi\,(\pl_{Y_1}\pl_{H_1}+\pl_{Y_2}\pl_{H_2})\mathcal{M}\,\xi_1(x_1,u_1)\,\xi_2(x_2,u_2)\Big|_{x_i=x,\,u_i=0}\,,\label{flatbr}
\end{equation}
where via integration by parts all derivatives are made to act on the gauge parameters, and $\Pi_\xi$ enforces tracelessness and the correct homogeneity degree in $x$. 

To determine the would-be higher-spin algebra structure constants, we evaluate \eqref{flatbr} on Killing tensors ${\bar \xi}$,
\begin{equation}
    u \cdot \partial_x\, {\bar \xi}\left(x,u\right) = 0. 
\end{equation}
In Minkowski space the Killing tensors are given by a set which transform as two-row Young tableaux:
\begin{equation}
   \xi^{(k)}(x,u)= \frac1{2^k(s-1)!k!}\,\xi_{\mu(s-1),\nu(k)}\,u^{\mu(s-1)}\,x^{\nu(k)}\,, \qquad k \leq s-1. \label{KillingFlat}
\end{equation}
The generalised higher-spin Lorentz generators correspond to those with $k=s-1$, while those with $k<s-1$ acquire a natural interpretation in terms of generalised hyper-translation generators. For simplicity we restrict to the former, where the higher-spin gravitational coupling should give rise to the structure constants of the type $f_{2ss}$. These specify the transformation properties of the spin-$s$ generators under the Lorentz part of the isometry. 

Owing to the inclusion of the ``additional'' two-derivative couplings, the following subtlety must be considered. Since these vertices are singular in covariant form, the deformed structure constants will involve terms which are non-polynomial in the $Y_i$ variables. This leads to singular expressions when considering a contraction with polynomial type functions, such as those in \eqref{KillingFlat}:
\begin{equation}\label{basicbracket}
     f_{s_1,s_2,s_3}^{(k)}\rightarrow [\![\xi_1,\xi_2]\!]_3^{(0)}=\frac{k}4\left[(s_1+s_2)G-(k-1)H_3Y_3\right]Y_1^{s_1-k}Y_2^{s_2-k}Y_3^{s_3-k}G^{k-2}\,.
\end{equation}
We adopt the prescription to simply drop the singular terms (which we justify below), and thus neglect them in the sequel. With this prescription we fix the bi-linear form as
\begin{align}
\kappa_{s,s^\prime}&=\left\langle\bar{\xi}^{(k)}_{s}|\bar{\xi}_{s^\prime}^{(k^\prime)}\right\rangle\\
&\equiv\delta_{s,s^\prime}\delta_{k,k^\prime}\,b_{s,k}\,\frac{(\pl_{u_1}\cdot\pl_{u_2})^{s-1}}{(s-1)!}\frac{(\pl_{x_1}\cdot\pl_{x_2})^{k}}{k!}\,\bar{\xi}_1^{(k)}(x_1,u_1)\,\bar{\xi}_2^{(k^\prime)}(x_2,u_2)\Big|_{x_i=0,u_i=0}\,,\nonumber
\end{align}
where the coefficients $b_{s,k}$ determined by requiring cyclicity of the corresponding $2$-$s$-$s$ structure constants. For the higher-spin Lorentz subalgebra (with $k=s-1$), these are 
\begin{equation}
b_{s,s-1}=(-1)^{s}\frac{\sqrt{\pi } \,  2^{8-2 s}}{\Gamma \left(s-\frac{1}{2}\right)}\,,\label{bss1}
\end{equation}
where, as for the AdS case in \S \tcb{\ref{subsec::hhasc}}, the overall constant has been fixed by normalising the $1$-$1$-$1$ structure constants to the identity. As a consistency check of our prescription for dealing with singular terms, \eqref{bss1} precisely reproduces the result obtained in the AdS$_4$ theory in \eqref{bcoeff} when normalising the kinetic term canonically. We also note that the bi-linear form is non-degenerate precisely due to the contribution of the lower-derivative exotic couplings.

The fact that from \eqref{basicbracket} we obtain the \emph{same} expression as for the $f_{2s_1s_2}$ structure constants \eqref{strconst} in AdS$_4$ suggests that the additional two-derivative $s$-$s$-2 vertices can be interpreted as minimal couplings of spin-$s$ gauge fields to gravity in flat space. Furthermore, this agreement suggests that the additional vertices we observe in the light-cone gauge should \emph{not} be considered as true independent additional vertices. Indeed, there exists a \emph{unique} combination of the standard local vertices and the exotic lower derivative ones:
\begin{equation}
    \mathcal{V}=\mathcal{V}_{\text{standard}}+\#\mathcal{V}_{\text{exotic}}\,,
\end{equation}
which admit an invariant bilinear form for the generalised Lorentz subalgebra of the putative higher-spin algebra. In AdS$_4$, due to the non-commutative nature of covariant derivatives, gauge invariance fixes such lower derivative vertices in combination with higher-derivative vertices. The fact that they appear to be independent vertices in flat space could be related to the singular nature of the flat-limit. Therefore, in flat space they look singular and they need to be added by hand, but the singularity disappears upon considering a gauge fixing to the light cone gauge in four-dimensions. For any given triplet of non-zero spins, we thus end up with one abelian higher-derivative vertex and one lower derivative cubic vertex which is a linear combination of standard local vertices and exotic ones (quasi-minimal coupling). The relative coefficients can be fixed by the requirement that the higher-spin Lorentz subalgebra admits an invariant bilinear form and the solution to quartic consistency precisely fulfils this requirement.

One can speculate that upon introducing auxiliary fields these vertices may be rewritten in local form. The analysis presented above may be interpreted as a hint that the corresponding theory has an underlying higher-spin symmetry, a possibility which we discuss further in the following section. 

\subsection{Is there a higher-spin algebra underlying Metsaev's vertices?}
In this section we give the extension of the result in the previous section for the $s$-$s$-$2$ structure constants to the generic case of $s_1$-$s_2$-$s_3$.

Since the computation is intrinsically four-dimensional, the following dimension dependent identity should be employed:
\begin{equation}
    M_{123}^2+\frac12 M_{12}M_{23}M_{31}=0\,,\label{DDI}
\end{equation}
which permits the removal of any power of $M_{123}$ greater than one. Accomodating for the above identity, we compute the full list of $4d$ structure constants induced by the cubic vertices \eqref{MetCov}: 
\begin{multline}\label{4dstrmet}
f^{(s_1,s_2,s_3)}=\sum_{l=0[1]}^{\text{Min}(s_1,s_2,s_3)-1}\bar{f}^{((s_1+s_2-s_3-l-1)/2,(s_2+s_3-s_1-l-1)/2,(s_3+s_1-s_2-l-1)/2,l)}\\\times\,M_{12}^{(s_1+s_2-s_3-l-1)/2}M_{23}^{(s_2+s_3-s_1-l-1)/2}M_{31}^{(s_3+s_1-s_2-l-1)/2}M_{123}^l\,,
\end{multline}
where $l$ ranges over the odd integers if $s_1+s_2+s_3$ is even, and over even integers if $s_1+s_2+s_3$ is odd. The coefficients ${\bar f}$ are defined by
{\footnotesize
\begin{align}
f^{(k_1,k_2,k_3,l)}&=\sum_{n=0}^\infty\sum_{m=0}^\infty \mathcal{P}_{m,n}^{(k_i,l)} \pi\, (-1)^m (k_1+k_2+l+1) (k_1+k_3+l+1) (k_2+k_3+l+1)\\
&\times\frac{2^{-3 (k_1+ k_2+ k_3)-4 l-2 m-6 n+5} \Gamma \left(m+2 n+\frac{1}{2}\right) \Gamma (m+2 n+2) \Gamma (2 k_1+2 k_2+2 k_3+3 l+3)}{\Gamma \left(m+3 n+2\right)\Gamma \left(k_1+k_2+l+\frac{1}{2}\right) \Gamma \left(k_1+k_3+l+\frac{1}{2}\right) \Gamma \left(k_2+k_3+l+\frac{1}{2}\right)}\,,\nonumber
\end{align}}
with
{\footnotesize
\begin{align}
\mathcal{P}_{m,n}^{(k_i,l)}&=\frac{1}{\Gamma (-k_1-k_2-l+m+2 n+1) \Gamma (-k_1-k_3-l+m+2 n+1) \Gamma (-k_2-k_3-l+m+2 n+1)}\nonumber\\
&\times\frac1{\Gamma (k_1+k_2+k_3+l-m-2 n+1) \Gamma (-k_1-k_2-k_3-l+m+3 n+1)}\,,\nonumber\\
&\times\frac1{\Gamma (2 k_1+2 k_2+2 k_3+3 l-2 m-6 n+1)}\,.
\end{align}}

\noindent This result precisely coincides with the same structure constants \eqref{strAdSd} in the AdS$_4$ theory, and thus illustrates that our formal covariantisation \eqref{MetCov} of the cubic vertices in \cite{Metsaev:1991mt,Metsaev:1991nb} uncovers the full Lorentz part of the higher-spin symmetry. The latter can also be rewritten in terms of the Moyal product in the enveloping algebra construction for $sl(2,\mathbb{C})$. We emphasise that the result \eqref{4dstrmet} crucially relies on lower-derivative exotic couplings, whose covariant form might require the addition of auxiliary field to be reduced to a standard local formulation.

The fact that the above higher-spin Lorentz structure constants precisely coincide with the AdS$_4$ higher-spin lorentz structure constants may also hint towards the existence of a well-defined relation between the theory in AdS$_4$ and in flat space. This is compatible with the existence of a contraction of the AdS$_4$ higher-spin algebra which naturally preserves its Lorentz part. In this respect it is important to make an analogous study of the hypertranslation-type global symmetries, which are not preserved in such a contraction. While we have not been able to determine the full list of structure constants for the hypertranslations, we have checked a number of lower spin examples. This would further clarify whether there exists an infinite dimensional extension of Poincar\'e algebra behind the cubic couplings \eqref{mcub}. Should it prove that the Metsaev theory is governed by a flat space higher-spin algebra, two main open questions then arise: First if the algebra advocated above can be realised as a contraction of the AdS$_4$ higher-spin algebra. Second, if there exists an oscillator realisation of it based on a universal enveloping algebra construction. Some attempts in this direction can be found in \cite{Bekaert:2008sa}, where some issues were also pointed out in looking for a proper way to factor the trace ideal. Some of the obstructions found in previous literature might be overcome via dimensional dependent identities, while they are expected to remain in $d > 4$.

\section{Spinor-helicity Formalism}\label{Sec:spinohelicity}

Recently there has been a renewed interest in the spinor-helicity formalism (see \cite{Dixon:2013uaa,Elvang:2013cua} for reviews on the subject) in the context of both massless \cite{Benincasa:2007xk} and massive \cite{Conde:2016vxs} higher-spins, with progress so far restricted to cubic amplitudes. Owing to the tight relation between light-cone and spinor helicity formalism (see for instance \cite{Ananth:2012un,Bengtsson:2016jfk} in the context of higher-spins), in this section we revisit this analysis in the light of our results presented in the preceding section. This is complementary to the recent work \cite{Conde:2016izb}, which studied the relation between cubic vertices in the original covariant classification (i.e. not accounting for the exotic lower derivative vertices of Metsaev considered in the present work) and three-point spinor helicity amplitudes.

A key feature of the spinor-helicity formalism is that 
the on-shell conditions for massless fields can be solved without giving up manifest covariance. For example, in this formalism the solution to the massless scalar Klein-Gordon equation is
\be
	\varphi(x)=
	\int \frac{d^{2}\lambda\, d^{2} \tilde \lambda}{{\text{vol}}(GL(1))}\,
	\exp\left(i\,\la\lambda|x^{\mu}\sigma_{\mu}|\tilde \lambda]\right)\,
	\phi(\lambda,\tilde \lambda)\,,
\ee
where we introduce two-component spinors $\lambda_a$ and ${\bar \lambda}_{\dot a}$, with \m{\la \lambda\,|\,\eta\ra=\lambda_{a}\,\eta_{b}\,\epsilon^{ab}} and\\ \m{[ \tilde \lambda\,|\,\tilde \eta]=
\tilde\lambda_{\dot a}\,\tilde\eta_{\dot b}\,\epsilon^{\dot a \dot b}}\,. Here, we have the following action of the little group on the polarisation tensors (see e.g. \cite{Benincasa:2007xk} and references therein for further details):
\be
	\phi(\Omega\,\lambda,\Omega^{-1}\,\tilde \lambda)
	=\phi(\lambda,\tilde \lambda)
	\qquad
	[\Omega \in \mathbb C]\,.	
\ee
In the following we review the generalisation of the above setting to higher-spin fields. For convenience, in four space-time dimensions one works with the fundamental representations $(1/2,0)$ and $(0,1/2)$ rather then with the usual vector oscillators. Hence we need to solve the following Fierz system:
\begin{eqnarray}
&&\square \, \varphi_{\alpha_1\ldots\alpha_s,\dot{\alpha}_1\ldots\dot\alpha_s}(x)\,=\,0\,\nn
&&\partial_{x}^{\,\alpha\dot\alpha}\,\varphi_{\alpha\ldots\alpha_s,\dot\alpha\ldots\dot\alpha_s}(x)\,=\,0\,,
\end{eqnarray}
while the traceless condition becomes automatic in this formalism due to the symmetrised form of the indices and to the antisymmetric nature of the $\epsilon_{\alpha\beta}$. As is standard in dealing with higher-spin fields, we introduce the generating function notation:
\be \varphi(x|\chi,\tilde\chi)=\sum_{s=0}^{\infty}\frac1{s!s!}\, \varphi_{\alpha_1\ldots\alpha_s,\dot\alpha_1\ldots\dot\alpha_s}(x) \, \chi^{\alpha_1}\,\cdots\,\chi^{\alpha_{s}}\,\chi^{\dot\alpha_1}\cdots\chi^{\dot\alpha_s}\,,
\ee
in terms of which the Fierz system reads
\be
	\Box\,\varphi(x|\chi,\tilde\chi)=0\,,\qquad
	\partial_{x}^{\,\alpha\dot\alpha}\partial_{\chi^\alpha}\partial_{\chi^{\dot\alpha}} \,\varphi(x|\chi,\tilde\chi)=0\,.
\ee
In this language the above equations can be solved in terms of a reference spinor which we denote by $q_\alpha$ and $\tilde q_{\dot\alpha}$. Going to momentum space and solving the mass-shell condition as
\be
p_{\alpha\dot\alpha}\,=\,\lambda_\alpha\,\tilde\lambda_{\dot\alpha}\,,
\ee
one gets the following general solution to the Fierz system:
\begin{multline}
\varphi(x|\chi,\tilde\chi)=\int \frac{d^{2}\lambda\, d^{2} \tilde \lambda}{{\rm vol}(GL(1))}\,
	\exp\left(i\,\la\lambda|x^{\mu}\sigma_{\mu}|\tilde \lambda]\right)\\\times
	\left[\left(\frac{\la\chi\lambda\ra\,[\tilde\chi\tilde q]}{[\tilde\lambda\tilde q]}\right)^s\phi_-(\lambda,\tilde \lambda)+\left(\frac{\la\chi q\ra\,[\tilde\chi\tilde\lambda]}{\la\lambda q\ra}\right)^s\phi_+(\lambda,\tilde \lambda)\right]\,.
\end{multline}
Let us note that the dependence on the auxiliary spinor is exactly compensated by the left-over on-shell gauge invariance; no dependence on the auxiliary spinor remains at the level of the amplitude. Indeed it is straightforward to prove that:
\be
\frac{\la\chi q^\prime\ra\,[\tilde\chi\tilde\lambda]}{\la\lambda q^\prime\ra}=\frac{\la\chi q\ra\,[\tilde\chi \tilde\lambda]}{\la\lambda q\ra}+\frac{\la q^\prime q\ra}{\la q^\prime\lambda\ra\la \lambda q\ra}\,\la\chi\lambda\ra\,[\tilde\lambda\tilde\chi]\,,
\ee
while the on-shell gauge invariance reads in this formalism:
\be
\delta_{\xi}\varphi(x|\chi,\tilde\chi)=\int \frac{d^{2}\lambda\, d^{2} \tilde \lambda}{{\rm vol}(GL(1))}\,
	\exp\left(i\,\la\lambda|x^{\mu}\sigma_{\mu}|\tilde \lambda]\right)
\la\chi\lambda\ra\,[\tilde\lambda\tilde\chi]\,\xi(\lambda,\tilde \lambda|\chi,\tilde\chi)\,.
\ee
The problem of writing couplings modulo field redefinitions can be then posed at the level of the fields
\be
\phi(\lambda,\tilde\lambda|\chi,\tilde\chi)\,=\,\left(\frac{\la\chi\lambda\ra\,[\tilde\chi\tilde q]}{[\tilde\lambda\tilde q]}\right)^s\phi_-(\lambda,\tilde \lambda)+\left(\frac{\la\chi q\ra\,[\tilde\chi\tilde\lambda]}{\la\lambda q\ra}\right)^s\phi_+(\lambda,\tilde \lambda)\,,
\ee
where the generic coupling has the form
\begin{multline}
\mathcal{V}^{(3)}\,=\,\delta\left(\sum_i\,\tilde\lambda_i\lambda_i\right)\,C(\lambda_i,\tilde\lambda_i,\partial_{\chi_i} ,\partial_{\tilde\chi_i})\\\times\,\phi_1(\lambda_1,\tilde\lambda_1|\chi_1,\tilde\chi_1) \,\phi_2(\lambda_2,\tilde\lambda_2|\chi_2,\tilde\chi_2)\,\phi_3(\lambda_3,\tilde\lambda_3|\chi_3,\tilde\chi_3)\Big|_{\chi_i=\tilde\chi_i=0}\,.
\end{multline}
A key point of the above expression is the $GL(1)$ invariance which must be imposed on the function $C$ together with gauge invariance.
In order to properly study the above problem it is useful to first determine the identities among the various $\lambda_i$ that are implied by momentum conservation.

To begin with, momentum conservation implies either
\be
\la 12\ra=\la23\ra=\la31\ra=0\,,
\ee
or
\be
[12]=[23]=[31]=0\,,
\ee
where $\la I J \ra = \lambda_{I,a}\,\lambda_{J,b}\,\epsilon^{ab}$ and $[ I J ]=
\bar \lambda_{I, \dot a}\,\bar\eta_{J,\dot b}\,\epsilon^{\dot a \dot b}$. The above precisely reduce to holomorphicity of the amplitudes recovered in the light-cone gauge. Furthermore, due to over-antisymmetrisation one also gains the identity
\be
\la I J\ra\la K L \ra+\la J K\ra\la I L\ra+\la K I\ra\la J L\ra\,=\,0\,.
\ee
At this point the possible building blocks are then given by
\be
\la \lambda_i\lambda_j\ra\,,\qquad \la\lambda_i\partial_{\chi_j}\ra\,,\qquad \la\partial_{\chi_i}\partial_{\chi_j}\ra\,,\qquad[ \tilde\lambda_i\tilde\lambda_j]\,,\qquad [\tilde\lambda_i\partial_{\tilde\chi_j}]\,,\qquad [\partial_{\tilde\chi_i}\partial_{\tilde\chi_j}]\,.
\ee
In the case that the $\tilde\lambda$ are proportional to each other, momentum conservation further implies 
\be
\lambda_1+\tfrac{\la 31\ra}{\la 23\ra}\,\lambda_2+\tfrac{\la 12\ra}{\la 23\ra}\,\lambda_3\,=\,0\,,
\ee
and similarly for the anti-holomorphic components.
This identity, together with the divergenceless condition and GL(1) invariance, reduces the number of independent building blocks to the following:
\ba
&&P_3=\la12\ra\,[12]\,,\quad P_1=\la23\ra\,[23]\,,\quad P_2=\la31\ra\,[31]\,,\nn&&Y_2=\la1\partial_{\chi_2}\ra\,[1\partial_{\tilde\chi_2}]\,,\quad Y_3=\la2\partial_{\chi_3}\ra\,[2\partial_{\tilde\chi_3}]\,,\quad Y_1=\la3\partial_{\chi_1}\ra\,[3\partial_{\tilde\chi_1}]\,,
\ea
while any other building blocks can be expressed in terms of the above modulo Fierz identities.
Making use of the following useful identities valid for any reference momentum $q$:
\ba
\frac{[3q]}{[1q]}\,=\,\frac{\la 12\ra}{\la 23\ra}\,,\quad \frac{[1q]}{[2q]}\,=\,\frac{\la 23\ra}{\la 31\ra}\,,\quad \frac{[2q]}{[3q]}\,=\,\frac{\la 31\ra}{\la 12\ra}\,,
\ea
together with their anti-holomorphic counterparts, we can now construct the couplings, classifying them depending on the helicity involved: i.e. $(+++)$, $(++-)$, $(+--)$ and $(---)$. One observes that for each helicity combination only one particular function $C$ gives a non-vanishing result.
\begin{itemize}
\item $(+++)$: In this case the non vanishing coupling is recovered from
\ba
C&=&Y_1^{s_1}\,Y_2^{s_2}\,Y_3^{s_3}\,\rightarrow\,\left(\frac{\la3q_1\ra[31]}{\la1q_1\ra}\right)^{s_1}\left(\frac{\la1q_2\ra[12]}{\la2q_2\ra}\right)^{s_2} \left(\frac{\la2q_3\ra[23]}{\la3q_3\ra}\right)^{s_3}\,\phi_{1+}\,\phi_{2+}\,\phi_{3+}\nn
&&=\left(\frac{[12][31]}{[23]}\right)^{s_1}\left(\frac{[23][12]}{[31]}\right)^{s_2} \left(\frac{[31][23]}{[12]}\right)^{s_3}\,\phi_{1+}\,\phi_{2+}\,\phi_{3+}\nn
&&=[12]^{s_1+s_2-s_3}[23]^{s_2+s_3-s_1}[31]^{s_3+s_1-s_2}\,\phi_{1+}\,\phi_{2+}\,\phi_{3+}\nn
&&=[12]^{h_1+h_2-h_3}[23]^{h_2+h_3-h_1}[31]^{h_3+h_1-h_2}\,\phi_{1+}\,\phi_{2+}\,\phi_{3+}\,.
\ea
\item $(++-)$: 
\ba
C&=&\left(\tfrac{P_3}{P_1P_2}\right)^{s_3}\,Y_1^{s_1}\,Y_2^{s_2}\,Y_3^{s_3}\,\nn&&\rightarrow\left(\tfrac{P_3}{P_1P_2}\right)^{s_3}\left(\frac{\la3q_1\ra[31]}{\la1q_1\ra}\right)^{s_1}\left(\frac{\la1q_2\ra[12]}{\la2q_2\ra}\right)^{s_2} \left(\frac{\la23\ra[2q_3]}{[3q_3]}\right)^{s_3}\,\phi_{1+}\,\phi_{2+}\,\phi_{3-}\nn
&&=\left(\frac{[12][31]}{[23]}\right)^{s_1}\left(\frac{[23][12]}{[31]}\right)^{s_2} \left(\frac{[12]}{[31][23]}\right)^{s_3}\,\phi_{1+}\,\phi_{2+}\,\phi_{3-}\nn
&&=[12]^{s_1+s_2+s_3}[23]^{s_2-s_3-s_1}[31]^{-s_3+s_1-s_2}\,\phi_{1+}\,\phi_{2+}\,\phi_{3-}\nn
&&=[12]^{h_1+h_2-h_3}[23]^{h_2+h_3-h_1}[31]^{h_3+h_1-h_2}\,\phi_{1+}\,\phi_{2+}\,\phi_{3-}\,. \label{++-}
\ea
\item $(+--)$: 
\ba
C&=&\left(\tfrac{P_1}{P_2P_3}\right)^{s_1}\,Y_1^{s_1}\,Y_2^{s_2}\,Y_3^{s_3}\,\nn&&\rightarrow\left(\tfrac{P_1}{P_2P_3}\right)^{s_1} \left(\frac{\la3q_1\ra[31]}{\la1q_1\ra}\right)^{s_1}\left(\frac{\la12\ra[1q_2]}{[2q_2]}\right)^{s_2} \left(\frac{\la23\ra[2q_3]}{[3q_3]}\right)^{s_3}\,\phi_{1+}\,\phi_{2-}\,\phi_{3-}\nn
&&=\left(\frac{\la23\ra}{\la12\ra\la31\ra}\right)^{s_1}\left(\frac{\la12\ra\la23\ra}{\la31\ra}\right)^{s_2} \left(\frac{\la23\ra\la 31\ra}{\la12\ra}\right)^{s_3}\,\phi_{1+}\,\phi_{2-}\,\phi_{3-}\nn
&&=\la12\ra^{-s_1+s_2-s_3}\la23\ra^{s_2+s_3+s_1}\la31\ra^{s_3-s_1-s_2}\,\phi_{1+}\,\phi_{2-}\,\phi_{3-}\nn
&&=\la12\ra^{-h_1-h_2+h_3}\la23\ra^{-h_2-h_3+h_1}\la31\ra^{-h_3-h_1+h_2}\,\phi_{1+}\,\phi_{2-}\,\phi_{3-}\,.\label{+--}
\ea
\item $(---)$: 
\ba
C&=&Y_1^{s_1}\,Y_2^{s_2}\,Y_3^{s_3}\,\rightarrow \left(\frac{\la31\ra[3q_1]}{[1q_1]}\right)^{s_1}\left(\frac{\la12\ra[1q_2]}{[2q_2]}\right)^{s_2} \left(\frac{\la23\ra[2q_3]}{[3q_3]}\right)^{s_3}\,\phi_{1-}\,\phi_{2-}\,\phi_{3-}\nn
&&=\left(\frac{\la31\ra\la12\ra}{\la23\ra}\right)^{s_1}\left(\frac{\la12\ra\la23\ra}{\la31\ra}\right)^{s_2} \left(\frac{\la23\ra\la 31\ra}{\la12\ra}\right)^{s_3}\,\phi_{1-}\,\phi_{2-}\,\phi_{3-}\nn
&&=\la12\ra^{s_1+s_2-s_3}\la23\ra^{s_2+s_3-s_1}\la31\ra^{s_3+s_1-s_2}\,\phi_{1-}\,\phi_{2-}\,\phi_{3-}\nn
&&=\la12\ra^{-h_1-h_2+h_3}\la23\ra^{-h_2-h_3+h_1}\la31\ra^{-h_3-h_1+h_2}\,\phi_{1-}\,\phi_{2-}\,\phi_{3-}\,.
\ea
\end{itemize}
While in restricting attention to the above building blocks gauge invariance is manifest, the above results are in complete agreement with those found in the previous sections in the light-cone gauge.\footnote{To see this one needs to employ the identity \be \left(\tfrac{P_3}{P_1P_2}\right) Y_1Y_2Y_3+\text{cycl.}\sim G\,,
\ee specific to four-dimensions. This follows from
\be
\frac{\la 12\ra\la3\partial_{\chi_1}\ra\la1\partial_{\chi_2}\ra\la2\partial_{\chi_3}\ra}{\la23\ra\la31\ra}=\frac{\la 31\ra\la2\partial_{\chi_1}\ra\la3\partial_{\chi_2}\ra\la2\partial_{\chi_3}\ra}{\la23\ra\la31\ra}=-\la\partial_{\chi_1}\partial_{\chi_2}\ra\la2\partial_{\chi_3}\ra.
\ee
} In more detail, using the dictionary \S \tcb{\ref{gaugevertex}} to go from covariant cubic structures to the light-cone, combined with the above we can go straight from light-cone gauge to the spinor-helicity formalism. This is a one-to-one map, and thus resolves the puzzle regarding the mis-match between the original covariant classification of cubic vertices and three-point S-matrix structures reviewed in \cite{Conde:2016izb}. Notice in particular the explicit non-local form of the vertices \eqref{+--} and \eqref{++-} involving one opposite helicity, as observed for the covariant counter-parts of the exotic light-cone vertices in \S \tcb{\ref{subsec::covmetv}}. 

Let us stress that in \cite{Conde:2016izb} it was observed that in mapping higher-derivative local higher-spin couplings to the spinor-helicity formalism 
lower derivative structures appear as total derivatives, up to terms proportional to the equations of motion. This is to be expected, for in the ambient space formalism the minimal coupling is precisely generated when considering the radial reduction of such total derivatives \cite{Joung:2011ww}. On the other hand, in this work we point out the existence of  additional lower-derivative couplings which are non-singular only in four-dimensions and which reproduce the spinor-helicity structures without being multiplied by vanishing factors as in the case of \cite{Conde:2016izb}. The price to pay is a mild non-local form of the corresponding covariant expressions for the vertices. Furthermore, the corresponding functional class has been argued (\S \tcb{\ref{subsec::lcl}}) to be fully compatible with the existence of non-trivial couplings avoiding the triviality argument of \cite{Barnich:1993vg}.


\section{Higher-spin algebras and higher-order amplitudes}\label{quarticorder}
It is illuminating to study in more detail the consequences beyond cubic order of a possible higher-spin symmetry (i.e. in the case that the structure constants \eqref{4dstrmet} yield a well-defined higher-spin algebra) behind Metsaev's cubic couplings. Similar investigations have been made in \cite{Weinberg:1980kq,Taronna:2011kt,Joung:2015eny}. Likewise, it turns out that the higher-spin symmetry places very strong constraints on the momentum dependence of any 4pt amplitude.\footnote{See also \cite{Bekaert:2009ud,Taronna:2011kt,Dempster:2012vw,Ponomarev:2016jqk} for studies of four-point amplitudes of higher-spin theories in flat space.} 

We study the action of a hypertranslation on a higher-spin field, as obtained from the cubic couplings \eqref{MetCov} extending the discussion of \cite{Joung:2015eny} to the Metsaev case. We first consider the deformation generated by the $0$-$r_1$-$r_2$ coupling with $r_1+r_2$ derivatives,\footnote{In this section we use $w$ for auxiliary vectors and $r$ to denote the spin, in order to avoid any confusion with the Mandelstam variables $u$ and $s$.} which possesses the standard covariant form 
\begin{equation}
    f_{r_1,0,r_3}^{(0)}=Y_1^{r_1}\,Y_3^{r_3},
\end{equation}
i.e. it does not originate from the additional exotic vertices, see \S \tcb{\ref{subsec::covmetv}}. The corresponding deformations of the gauge transformations for a spin-$r_3$ field are given by:
\begin{equation}
    \delta\varphi_{r_3}(w)=r_1\,Y_1^{r_1-1}Y_3^{r_3}\,\xi_1^{(0)}(w_1)\phi_2\Big|_{w_1=0}\,,
\end{equation}
which rotates the spin-$r_3$ field into a scalar $\phi_2$ through a hypertranslation $\xi^{(0)}_{r_1}(w)$.
The latter however vanish identically if $r_3>0$. This follows from the following identity for hypertranslations:
\begin{equation}
Y_3\,\xi_1^{(0)}=w\cdot\pl_1 \xi_1^{(0)}(w_1)=0\,.\label{IdHyp}
\end{equation}
In the absence of exotic couplings with lower derivatives, this has a very simple consequence: It implies that the four-scalar amplitude should rotate into itself under hypertranslations $\xi$,
\begin{equation}
    \delta_{\xi^{(0)}}\mathcal{A}_{0000}=0\,.\label{hyp0}
\end{equation}
In more detail, consider the following spin-$r$ hyper-translation transformations of a scalar field
\begin{equation}
    \delta_{\xi^{(0)}_r}\phi_i(p)=g_{i}\,\xi^{(0)}_{\mu(r-1)}(i\,p)^{\mu(r-1)}\,\phi_i(p)\,,
\end{equation}
where $g_i$ are the coupling constants entering the cubic action of the theory.
The condition \eqref{hyp0} requires
\begin{equation}
  \delta_{\xi^{(0)}_r}\mathcal{A}_{0000} =  \xi^{(0)}_{\mu(r-1)}\,\mathcal{A}_{0000}(s,t,u)\sum_{i}g_i\,(i\,p_i)^{\mu(r-1)}=0\,.
\end{equation}
The above identity is simply a different incarnation of Weinberg result \cite{Weinberg:1964ew}, but follows here as a consequence of higher-spin symmetry as opposed to the consideration of a soft limit for the external particles. In particular, given the arbitrariness of the gauge parameter $\xi$, this is equivalent to
\begin{equation}\label{offending}
   \mathcal{A}_{0000}(s,t,u) \sum_{i=0}^4g_i\,(p_i)_{\mu_1}\cdots (p_i)_{\mu_{r-1}}=0\,.
\end{equation}
For spins $r=1$ and $r=2$ this enforces charge conservation and equivalence principle. For $r>2$ and $g_i\neq0$ however, the it implies that the scalar amplitude itself must be a distribution concentrated on a measure zero set of kinematical configurations which allow the above identity to be satisfied. The amplitude must then be concentrated on kinematical configurations which solve
\begin{equation}
    \sum_{i=0}^4g_i\,(p_i)_{\mu_1}\cdots (p_i)_{\mu_{r-1}}=0\,.
\end{equation}
These are the configurations in which the particles do not interact and where least one of the Mandelstam variables vanishes:
\begin{equation}
    \mathcal{A}_{0000}=a(s,t)\,\delta(u)+a(t,u)\,\delta(s)+a(u,s)\,\delta(t)\,,\label{trivial}
\end{equation}
with
\begin{align}
s&=-(p_1+p_2)^2\,,& t&=-(p_1+p_3)^2\,,& u&=-(p_1+p_4)^2\,,
\end{align}
Indeed, for instance $u=0$ implies that $t=-s$ and $p_1=-p_4$ and $p_2=-p_3$, i.e. triviality, together with analogous results for other channels. In this illustrative $u=0$ example we end up with
\begin{multline}
    \sum_{i=0}^4g_i\,(b\cdot p_i)^{r-1}= \Big[(-1)^{r-1}\,g_1\,(b\cdot p_4)^{r-1}\\+(-1)^{r-1}\,g_2\,(b\cdot p_3)^{r-1}+g_3\,(b\cdot p_3)^{r-1}+g_4\,(b\cdot p_4)^{r-1}\Big]=0\,,
\end{multline}
where $b$ is an arbitrary vector. This is satisfied if the $g_i$ are equal and their colored/charged legs satisfy the appropriate antisymmetry conditions for odd spins. We thus see that higher-spin symmetry forces the theory to have trivial $S$-matrix at quartic order, and that the standard Weinberg result is recovered from higher-spin symmetry.

We now re-consider the above discussion, but this time including the effect of the exotic couplings which are present in the cubic Lagrangian \eqref{MetCov}. Exotic coupling may indeed provide a way out to the above argument. Since $Y_3$ annihilates the hypertranslation generators, it is sufficient to consider the deformation to gauge transformations coming from $0$-$r_1$-$r_2$ exotic cubic couplings with no $Y_3$ dependence, namely
\begin{equation}
    \delta\varphi_{r_3}(u)=\,g_{r_1,0,r_3}\,(r_1-r_3)\,Y_1^{r_1-r_3-1}\left(H_2+\ldots\right)^{r_3}\,\xi_1^{(0)}(u_1)\,\phi_2\Big|_{u_1=0}\,,\label{exohyp}
\end{equation}
where the $\ldots$ are singular terms proportional to $Y_2^{-1}$, which are non-singular upon gauge fixing and do not contribute on global symmetries. Since in this case the transformation \eqref{exohyp} is non-vanishing, the four-scalar amplitude should not be invariant under higher-spin hypertranslations as in \eqref{hyp0}. By global higher-spin symmetry, it must be compensated by the variation of  amplitudes involving a single spinning external leg, where the transformation acts on the latter external field:
\begin{equation}
    \delta_{\xi^{(0)}_r}\mathcal{A}_{0000}+ \sum^\infty_{k=1}\delta_{\xi^{(0)}_r} \mathcal{A}_{k000} = 0.
\end{equation}
Assuming that there is no $k000$ exotic structure like for the $k00$ case, the most general form for a planar $k$-$0$-$0$-$0$ amplitude in the $s$ and $u$ channels compatible with Poincar\'e invariance reads \cite{Taronna:2011kt}:\footnote{We restrict our attention to amplitudes which can be dressed with Chan-Paton factors owing to the fact that Metsaev's solution \eqref{mcub} admits such an extension. The case without Chan-Paton factors is slightly more general, but the conclusions presented below for the case with Chan-Paton factors continue to hold.}
\begin{equation}\label{r000}
    \mathcal{A}_{k000}=\frac{f_k(t)}{su}\,\frac1{k!}\,\left(u\,\pl_{w_1}\cdot p_2-s\,\pl_{w_1}\cdot p_4\right)^k\,\phi_1(w_1,p_1)\phi_2(p_2)\phi_3(p_3)\phi_4(p_4)\Big|_{w_1=0}\,,
\end{equation}
with $f_k$ an arbitrary function of one variable and with no pole in the complex plane.\footnote{This assumption follows from factorisation and unitarity.} The hypertranslation transformation of this amplitude reads
\begin{equation}\label{HSvar}
   \delta_{\xi^{(0)}_r} \mathcal{A}_{k000} = \xi^{(0)}_{\mu(k)\nu(r-k-1)}g_{r,0,k}f_k(t)\left(u\,p_2-s\,p_4\right)^{\mu(k)}p_1^{\nu(r-k-1)}\,,
\end{equation}
where we sum over the action of the symmetry transformation on all external legs and without loss of generality restrict to the part of the variation which generates a 4-scalar structure (as the other structures are related to this one by higher-spin symmetry). The assumption that $f_k(t)$ does not contain poles in $t$ then implies that the contributions with $k>0$ in \eqref{HSvar} contain a number of derivatives $N_k$ which is bounded from below $N_k\geq 2k+r-1$. The contribution with $r$ derivatives (those in \eqref{HSvar} for $k=0$) thus cannot be compensated by any other amplitude. One can then argue that the only way to obtain an amplitude which is consistent with higher-spin symmetry is to have
\begin{equation}
    f_k(t) \: \sim \: \delta(t)\,,
\end{equation}
extending the previous result \eqref{trivial}. Notice that the above conditions only arise for amplitudes with the number of external legs being greater than or equal to four. At cubic level there is no non-trivial Mandelstam invariant and non-trivial cubic couplings are compatible with higher-spin symmetries. By higher-spin covariance, this result suggests that any 4-point amplitude is proportional to a sum of $\delta$-function distributions, reminiscent of AdS Mellin-amplitudes that can be extracted from free theories \cite{Taronna:2016ats,Bekaert:2016ezc}.

We emphasise that a key assumption of the above discussion is the absence of $r$-$0$-$0$-$0$ exotic structures. This is motivated by the absence of $r$-$0$-$0$ exotic structures, but should be verified.

\section{Conclusions}\label{Conclusions}
In this paper we studied the higher-spin algebra structure constants induced by the action at cubic order of the type A minimal bosonic higher-spin theory on AdS$_{d+1}$ space implied by holography \cite{Sleight:2016dba}. The explicit form of the structure constants for the deformed gauge symmetries are obtained, together with the associated normalisation of the bilinear form. We show that these structure constants coincide with the known expressions, which are unique in generic dimensions. This demonstrates that the holographically reconstructed cubic action solves the Noether procedure up to quartic order, and extends the tree-level three-point function test of the higher-spin / vector model duality by Giombi and Yin in AdS$_4$ \cite{Giombi:2009wh} to general dimensions.

We also considered the same problem for the cubic order action of the theory in 4d flat space found by Metsaev in \cite{Metsaev:1991mt,Metsaev:1991nb}. The couplings themselves where obtained by solving the quartic consistency in the light-cone formulation, where higher-spin symmetry is not manifest. Remarkably, the couplings include lower derivative vertices which were not captured in the original covariant classification of cubic structures \cite{Boulanger:2008tg,Zinoviev:2008ck,Manvelyan:2010jr,Sagnotti:2010jt,Fotopoulos:2010ay}. These include  two-derivative couplings of higher-spin fields to gravity, which we argued to be minimal. After extracting the explicit form of the higher-spin structure constants, we argue in favour of a well-defined higher-spin algebra behind the cubic couplings. The existence of such a higher-spin algebra crucially relies on the additional couplings couplings in flat 4d Minkowski space, initially found in \cite{Bengtsson:1983pd,Bengtsson:1986kh}. 

We end with a few summarising remarks and outlooks:
\begin{itemize}
\item Extending the dicussion of \S \tcb{\ref{quarticorder}}, there are indeed various examples in the literature where the implications of higher-spin symmetry on higher-order amplitudes have been considered. To date there are compelling arguments that both conformal higher-spin theories in flat space \cite{Joung:2015eny,Beccaria:2016syk} and higher-spin theories in AdS \cite{Taronna:2016ats,Bekaert:2016ezc}\footnote{The the context of the AdS/CFT duality, the analogue of the S-matrix in AdS has been argued to be the Mellin transform of the dual CFT correlators \cite{Mack:2009mi,Mack:2009gy,Penedones:2010ue,Fitzpatrick:2011ia,Paulos:2011ie}. The Mellin transform of correlation functions in a free CFT are ill-defined, since they are power-functions in the cross-ratios. However they can be formally defined as a $\delta$-function distribution \cite{mellin}.} have trivial S-matrix-like observables. These have been shown to be proportional to delta-function-like distributions concentrated on measure zero space for kinematic configurations. Together with the same story in flat space \S \tcb{\ref{quarticorder}}, this seems to point towards higher-spin symmetry being incompatible with a non-trivial S-matrix, at least in all known examples. 
\item It would be interesting explore the possibility of other covariantisations of the exotic light-cone vertices, which may avoid the formal singularities obtained through the covariantisation prescription in this note. It is conceivable that this would only be possible upon introducing infinitely many auxiliary fields.
\item Finally, another interesting direction would be to check the results obtained in this note directly in the light-cone gauge, skipping the step of covariantisation of the vertices. Some ideas in this direction have been discussed in \cite{Bengtsson:2012dw}. Indeed the analysis carried out in this work demonstrates that a well defined formulation of the exotic vertices is only available so far in a Lorentz non-covariant frame. It would also be interesting to investigate the relations between this case and the case of self-dual forms where a similar covariantisation problem arises \cite{Pasti:1995tn}.
\end{itemize}

\section*{Acknowledgments}
\label{sec:Aknowledgements}

We are indebted to G. Barnich, X. Bekaert and R. Metsaev for useful discussions and comments. We thank D. Ponomarev and E. Skvortsov for communications after v1 of this manuscript appeared, kindly letting us know about the unpublished work behind their talk at the ``Higher-spin Theory
and Duality" conference at MIAPP, Munich, 23-25, May 2016. The results of which then later appeared in \cite{Ponomarev:2016lrm}. The research of M. Taronna is partially supported by the Fund for Scientific Research-FNRS Belgium, grant FC 6369 and by the Russian Science Foundation grant 14-42-00047 in association with Lebedev Physical Institute. This research was also supported by the Munich Institute for Astro- and Particle Physics (MIAPP) of the DFG cluster of excellence Origin and Structure of the Universe.

\begin{appendix}

\section{Higher-spin algebra structure constants}\label{appendix}

Higher-spin algebra structure constants for symmetric tensor fields can all be obtained via universal enveloping algebra constructions \cite{Vasiliev:2004cm,Joung:2014qya}:\footnote{In this appendix $d$ refers to the space-time dimension of the higher-spin theory and not the dimension of the dual CFT.}
\be
hs(\mathcal{I})=\frac{\mathcal{U}(so(d,2))}{\mathcal{I}}\,.\label{quotient}
\ee
The above is usually realised through the use of oscillators by embedding the isometry algebra of AdS space into $sp(2N)$ for appropriate choices of $N$.
For all known higher-spin algebras involving totally-symmetric fields, the quotient operation is conveniently realised by a quasi-projector $\Delta$, \cite{Vasiliev:2004cm,Joung:2014qya}. The quasi-projector $\Delta$ can be defined as a non-polynomial element of the universal enveloping algebra of $sp(2N)$ which by construction projects out the ideal and picks a well-defined representative in \eqref{quotient}. Below we summarise the list of known higher-spin algebras in various dimensions and for totally symmetric fields. Apart from one parameter families arising in $d=3$ and $d=5$, the higher-spin algebra for totally symmetric tensors is unique.

\paragraph{The metaplectic representation} The starting point to treat all higher-spin algebras in Table~\ref{Table:HSalg} in a unified fashion is given by the metaplectic representation of $sp(2N)$ which is defined by:
\be
[Y_\gA,Y_\gB]_{\star}=2i C_{\gA\gB}\,,
\ee
with $C_{\gA\gB}$ the $sp$-invariant tensor, $C_{\gA\gC}C^{\gB\gC}=\delta^\gB_\gA$, which is used to raise and lower the indices according to $Q^{\gA}=C^{\gA\gB}Q_{\gB}$.
Above we introduced the Moyal $\star$-product in the Weyl-ordering, whose integral representation is
\be
(f\star g)(Y)=\int d^{2N}U d^{2N}V f(Y+U) g(Y+V)\,e^{iC^{\gA\gB}U_{\gA} V^{\gB}}\,,
\ee
so that discarding boundary terms one recovers
\begin{align}
Y_\gA\star&=Y_\gA+ i\pl_{\gA}& \star Y_\gA&=Y_\gA- i\pl_{\gA}\,,
\end{align}
with
\be
\pl_{\gA}Y_{\gB}\equiv C_{\gA\gB}\,.
\ee
The $sp(2N)$ generators read
\be
T_{\gA\gB}=-\frac{i}{2}\,Y_\gA Y_\gB\,,
\ee
and following \cite{Joung:2014qya}, it is convenient to introduce Gaussian generating functions for universal enveloping algebra elements of the type 
\be
g=e^{-\tfrac12 T_{\gA\gB}U^{\gA\gB}}\,.\label{spelem}
\ee
Above, the auxiliary variables $U^{\gA\gB}$ are assumed to be factorisable: $U^{\gA\gB}=u^{\gA}u^{\gB}$, so that $U^{\gA[\gB}U^{\gC]\gD}=0$ and the tracelessness of the generators is automatic $U_\gA{}^\gB U_{\gB}{}^\gC=0$. Finally, the above oscillator realisation admits a unique supertrace \cite{Vasiliev:1999ba}:
\be
\text{Tr}(f(Y))=f(0)\,.
\ee

\begin{table}
\begin{center}
{\small
\begin{tabular}{|c|c|c|c|c|c|}
\hline
Dim. & $AdS$-algebra & Howe-dual pair & Oscillators & $\mathcal{I}$\\
\hline
$d=3$ & $sl(2,\mathbb{R})\oplus sl(2,\mathbb{R})$ & $\left(gl(1),gl(2)\right)\in sp(4)$ & $\begin{matrix}
Y_\gA=(y^+_\ga,y^{-\ga})\\[3pt]
\phi,\ (\phi^2=1)
\end{matrix}$
& $y^{-\ga}y^+_{\ga}-\lambda\sim 0$\\[3pt]
\hline
d=4 & $sp(4,\mathbb{R})$ & $sp(4)$ & $
Y_\gA=(y_\ga,\bry_{\gad})$
&\\[3pt]
\hline
$d=5$ & $su(2,2)$ & $\left(gl(1),gl(4)\right)\in sp(8)$ & $\begin{matrix}
Y_\gA=(y^+_\ga,y^{-\ga})
\end{matrix}$
& $y^{-\ga}y^+_{\ga}-2\lambda\sim 0$\\[3pt]
\hline
$d\geq6$ & $so(d,2)$ & $\left(sp(2),so(d,2)\right)\in sp(2d+2)$ & $Y_A=y_{i\,a}$
& $\eta^{ab}y_{i\,a}y_{j\,b}\sim0$\\[3pt]
\hline
\end{tabular}}
\end{center}
\caption{Higher-spin algebras in various dimensions.}
\label{Table:HSalg}
\end{table}

\paragraph{The $\star$-monoid} The oscillator realisation introduced above enables the structure constants of the various higher-spin algebras to be encoded in terms of the unique $star$-product and trace operation defined in $sp(2N)$.

To this effect it is useful to recall the Monoid structure of Gaussians under the star product, which reproduces the $sp$-product up to a Caley transform \cite{Didenko:2003aa}
\be
F(S_1)\star F(S_2)=F(S_1S_2)\leftrightarrow F(S)=\frac1{\sqrt{\text{det}_{2N}\left(\frac{1+S}{2}\right)}}\,e^{\tfrac{i}{2}Y_\gA\left(\frac{S-1}{S+1}\right)^{\gA\gB}Y_{\gB}}\,,
\ee
where by definition $S\equiv S_\gA{}^\gB$ and $S_1S_2$ is the standard matrix multiplication $S_{1}{}_{\gA}{}^{\gB}S_2{}_{\gB}{}^{\gC}$ (recall that indices are raised and lowered with the invariant $sp$-tensor $C_{\gA\gB}$).

The above monoid structure allows multiple $\star$-products to be evaluated without the need for any computation, except for simply inverting the Caley transform
\be
S=\frac{1+\tfrac{U}{2}}{1-\tfrac{U}{2}}\,.
\ee
For example:
\be\label{strHS}
\text{Tr}\left(g_1\star\ldots\star g_k\right)=\left(\Det_{2N}\left[\frac12\prod_{i=0}^k(1+U_i)+\frac12\right]\right)^{-\tfrac12}\,.
\ee

\paragraph{Quasi-projectors} It is straightforward to restrict the universal enveloping algebra to its subalgebras. In the $so(d,2)$ case this simply amounts to the following choice of generators:
\begin{align}
    &e^{y^-_a\,y^+_b\,W^{ab}}\,,& W^{ab}&=w^{[a}_+w^{b]}_-\,.
\end{align}
However in doing so ideals emerge, which are generated by trace components. The simplest way to factor them out is to change the definition of the trace while keeping the same oscillator realisation induced by $sp(2N)$. This can be achieved by the following ansatz for the trace on the respective quotient \cite{Vasiliev:2004cm,Joung:2014qya}:
\be
\Tr_{g}[f(Y)]=(\Delta_g\star f)\Big|_{Y=0}\,.
\ee
The element $\Delta_g$ is defined by the requirement that it fixes a representative in the $\star$-product algebra. 

Given the Howe dual pair $(sp(2),so(d-1,2))$ and considering the trivial representation for the $sp(2)$, we have the ideal $y_{i}\cdot y_{j}\sim 0$. This ideal can be quotiented by the following quasi-projector \cite{Joung:2014qya}:
\be
\Delta_{so(N)}=\frac{\Gamma(\tfrac{N-1}{2})}{\Gamma(\tfrac{3}2)\Gamma(\tfrac{N-4}2)}\,\int_0^1 dx\,x^{\frac{N-6}2} (1-x)^{\tfrac12} e^{-2\sqrt{1-x}\,y^+\cdot y^-}\,.
\ee
The above Gaussian structure of the quasi-projector (first obtained in \cite{Vasiliev:2004cm}, but in a different form), makes it possible to extract all structure constants for the respective higher-spin algebras from the corresponding $sp(2N)$ structure constants.

\paragraph{$hs[so(N)]$ structure constants}
Evaluating the determinant \eqref{strHS} (see \cite{Joung:2014qya}) gives:
\begin{subequations}
\begin{align}
\Tr\left[e^{2\rho \,y^{-\,\ga}y^+_\ga}\star e^{ \tfrac12 T_{ab}W_1^{ab}}\star e^{ \tfrac12 T_{ab}W_2^{ab}}\right]&=\left[1+\frac{1-\rho^2}8 \,\Tr(W_1W_2)\right]^{-2}\,,\\
\Tr\left[e^{2\rho \,y^{-\,\ga}y^+_\ga}\star e^{ \tfrac12 T_{ab}W_1^{ab}}\star e^{ \tfrac12 T_{ab}W_2^{ab}}\star e^{ \tfrac12 T_{ab}W_3^{ab}}\right]&\\&\hspace{-40pt}=\left[\left(1+\frac{1-\rho^2}8 \Lambda\right)^2-\frac{\rho^2(1-\rho^2)^2}{32}\,\Sigma\right]^{-1}\,,\nonumber
\end{align}
\end{subequations}
with
\begin{subequations}
\begin{align}
\Lambda&=M_{12}+M_{23}+M_{31}+M_{123}\,,\\
\Sigma&=\frac12M_{123}^2+\frac14M_{12}M_{23}M_{31}\,,
\end{align}
\end{subequations}
in terms of the contractions of eq.~\eqref{Ms}. Evaluating the integrals in the quasi-projector then gives the following bilinear form:
\begin{equation}
    k(M_{12})=\sum_{i=0}^\infty(i+1)\,\frac{\left(\frac{N-4}{2}\right)_i}{\left(\frac{N-1}{2}\right)_i}\left(-\frac{1}{8}M_{12}\right)^{i}\,,
\end{equation}
together with the following structure constants:
\begin{multline}
f^{(s_1,s_2,s_3)}=\sum_{l=0[1]}^{\text{Min}(s_1,s_2,s_3)-1}\bar{f}^{((s_1+s_2-s_3-l-1)/2,(s_2+s_3-s_1-l-1)/2,(s_3+s_1-s_2-l-1)/2,l)}\\\times\,M_{12}^{(s_1+s_2-s_3-l-1)/2}M_{23}^{(s_2+s_3-s_1-l-1)/2}M_{31}^{(s_3+s_1-s_2-l-1)/2}M_{123}^l\,,
\end{multline}
where the sum over $l$ ranges over the odd integers if $s_1+s_2+s_3$ is even and over even integers if $s_1+s_2+s_3$ is odd. We have also defined the coefficients $\bar{f}^{(k_1,k_2,k_3,l)}$ as:

{\footnotesize
\begin{align}
\bar{f}^{(k_1,k_2,k_3,l)}&=\sum_{n=0}^\infty\sum_{m=0}^\infty \mathcal{P}_{m,n}^{(k_i,l)} \frac{(-1)^m2^{-k_1-k_2-k_3-l-2 m-6 n}\Gamma (m+2 n+2)\Gamma \left(\frac{N-1}{2}\right) \Gamma \left(m+2 n+\frac{N}{2}-2\right)}{\Gamma \left(\frac{N-4}{2}\right) \Gamma \left(m+\frac{1}{2} (6 n+N-1)\right)}\,,
\end{align}}

\noindent
\!with
{\footnotesize
\begin{align}
\mathcal{P}_{m,n}^{(k_i,l)}&=\frac{1}{\Gamma (m+2 n-l-k_1-k_2+1) \Gamma (m+2n-l-k_1-k_3+1) \Gamma (m+2n-l-k_2-k_3+1)}\nonumber\\
&\times\frac1{\Gamma (k_1+k_2+k_3+l-m-2 n+1) \Gamma (m+3n-l-k_1-k_2-k_3+1)}\,,\nonumber\\
&\times\frac1{\Gamma (2 k_1+2 k_2+2 k_3+3 l-2 m-6 n+1)}\,.
\end{align}}

\noindent
\!Upon normalising the bilinear form to the identity, the above form of the higher-spin algera structure constants coincides with the structure constants \eqref{strconst} extracted from the holographically reconstructed cubic vertices \eqref{full}.

\end{appendix}

\bibliography{megabib}
\bibliographystyle{JHEP}

\end{document}